%% file: main.tex
\documentclass[a4paper,11pt, notitlepage, oneside]{article} 
\usepackage[T1]{fontenc}
\usepackage[applemac]{inputenc}
\usepackage{lmodern} 
\usepackage[english]{babel}
\usepackage[titletoc,toc,title]{appendix}
\usepackage{titling}

\usepackage{graphicx,epsfig, color}
\usepackage[dvipsnames]{xcolor}

\usepackage{geometry}
\usepackage{fancyhdr}
\usepackage{setspace} 
\onehalfspace
\usepackage[multiple,bottom]{footmisc}
\usepackage{datetime} 
\newdateformat{monthyeardate}{%
  \monthname[\THEMONTH] \THEYEAR}
  \usepackage{setspace} %

\usepackage[autostyle]{csquotes}
\usepackage{natbib}
\setcitestyle{authoryear}
\usepackage{url} %
\usepackage{caption}%
\usepackage{subcaption}
\usepackage{capt-of}

\usepackage{amsmath,amssymb,amsthm,amsfonts,amsbsy}
\usepackage{mathrsfs}
\usepackage{libertine} 
\usepackage{bbm} 

\usepackage{adjustbox}
\usepackage{tabularx}
\usepackage{multicol}
\usepackage{multirow}
\usepackage{threeparttable} 
\usepackage{color, colortbl}
\usepackage[capposition=top]{floatrow}
\usepackage{lscape}
\usepackage{booktabs}
\usepackage{rotating}
\usepackage{float}
\usepackage{longtable, array, afterpage}
\usepackage{tablefootnote}   
\usepackage[colorlinks=true,linkcolor=RoyalBlue, citecolor=RoyalBlue, urlcolor=blue]{hyperref} 

\definecolor{Gray}{gray}{0.9}

\renewcommand{\Omega}{\varOmega}

\def\E {\mathbb{E}}

\def\F {\mathbb{F}}
\def\P {\mathbb{P}}
\newcommand{\Tau}{\mathrm{T}}

\newcommand{\indep}{\perp \!\!\! \perp}

\theoremstyle{plain}
\newtheorem{assumption}{Assumption}

\numberwithin{intassumption}{assumption}

\newtheorem{theorem}{Theorem}

\newtheorem{corollary}{Corollary}
\newtheorem{prop}{Proposition}

\date{May 21, 2025}

\title{Distributional Difference-in-Differences with Multiple Time Periods\thanks{This is a preliminary version of the paper where most of the comments and feedback received still need to be addressed. All the Monte Carlo simulations were run in STATA. The ado files of the command used for implementing the methodology presented in the paper, \texttt{qtt}, are available upon request at the time of writing.}}
\author{%
\textsc{Andrea Ciaccio}\thanks{\noindent Department of Economics, Ca' Foscari University of Venice, Italy. E-mail: \href{mailto:andrea.ciaccio@unive.it}{\textcolor{blue}{andrea.ciaccio@unive.it}}}
}

\begin{document}
\maketitle
\begin{abstract}
\noindent
Researchers are often interested in evaluating the impact of a policy on the entire (or specific parts of the) distribution of the outcome of interest. In this paper, I provide a method to recover the whole distribution of the untreated potential outcome for the treated group in non-experimental settings with staggered treatment adoption by generalizing the existing quantile treatment effects on the treated (QTT) estimator proposed by \cite{CallawayLi2019}. Besides the QTT, I consider different approaches that anonymously summarize the quantiles of the distribution of the outcome of interest (such as tests for stochastic dominance rankings) without relying on rank invariance assumptions. The finite-sample properties of the estimator proposed are analyzed via different Monte Carlo simulations. Despite being slightly biased for relatively small sample sizes, the proposed method's performance increases substantially when the sample size increases.

\bigskip

\textbf{Acknowledgments:} I am grateful to Manuel Arellano, Dmitry Arkhangelsky, Federico Belotti, Paolo Li Donni, Irene Mammi, Mario Padula, and Davide Raggi for their invaluable advice, insights, constructive suggestions, and detailed feedback on previous drafts of this paper. I also thank Francesco Chiocchio, Alexander Simon Mayer, and Annalivia Polselli for their thoughtful comments on an earlier version. Additionally, I appreciate the valuable feedback from participants at the CEMFI Econometrics Workshop and the Ca' Foscari Internal Departmental Seminar.
\end{abstract}

\clearpage
\input{Introduction.tex}
\input{Identification.tex}
\input{Estimation.tex}
\input{Simulation.tex}
\input{Conclusion.tex}
\newpage 
\bibliography{Bibliography.bib}

\clearpage
\appendix
\label{Appendix}
\input{AppendixA}
\clearpage
\input{Appendix_repeated}

\clearpage
\input{AppendixB}

\end{document}

%% file: Introduction.tex
\section{Introduction}
\label{Introduction}

In economics and, more generally, in social sciences, we are often interested in assessing the impact of a policy intervention on an outcome of interest in non-experimental settings. However, researchers face a selection problem when assessing the policy's causal effect. Once the policy is implemented, data will not reveal counterfactual outcomes (i.e., the outcomes one would have observed had the policy not been implemented) \citep{Holland1986}. The researcher's main challenge will be finding an appropriate way to recover the missing counterfactual outcomes.

Different methods exist to estimate the causal effect of policy interventions in non-experimental settings. One of the most popular methods is the \textit{Difference-in-Differences (DiD)}. In the classical DiD setting, we have two periods and two groups: in the first period, no unit is treated; in the second period, some units become treated (treated group) while others remain untreated (untreated group). Under the assumption that had the policy not been implemented, the average evolution between treated and untreated units would have been parallel (also known as \textit{common trends} or \textit{parallel trends (PT)} assumption), it is possible to estimate the (average) counterfactual outcome for the treated group consistently. Once the counterfactual outcome is retrieved, the most common treatment effect parameter considered in these settings is the Average Treatment Effect on the Treated (ATT), which estimates the average causal effect of the policy for the treated subpopulation.\footnote{The DiD's popularity is mainly due to its broad applicability to many research questions and the fact that, if we have many independent clusters, one can prove that we can consistently estimate the ATT using a two-way fixed effects (TWFE) regression.} Specifically, estimation of the ATT is achieved by comparing the average change in the outcomes observed for the treated subpopulation to that experienced by the untreated group.

While most studies employ causal inference methods to estimate the ATT, sometimes researchers might also be interested in evaluating the impact of a policy on the entire (or specific parts of the) distribution of the outcome of interest. For instance, think of two policies that aim at increasing wages but with the same average impact. Policymakers will prefer the policy that is more likely to lead to a more significant increase in the lower deciles of the income distribution than the one that should generate higher (expected) benefits for individuals "dwelling" in the middle-top deciles. Similarly, there are plenty of situations in economics where considering the entire distributional effect of a policy is more appropriate, especially when there are reasons to expect the impact to be heterogeneous across treated groups. 

One way to examine the distributional effect of a policy is to consider the Quantile Treatment Effect on the Treated (QTT). The QTT estimates the causal effect of a policy -- for the treated group -- on a specific quantile of the outcome of interest by comparing the quantiles of treated and untreated outcomes. Different studies exist that show how to retrieve a consistent estimator of the QTT in two-period two-group settings \citep{AtheyImb2006, BonhommeSauder, Fan2012, Callawayetal2018, CallawayLi2019, Miller2023}. 

In this paper, I propose a method to recover counterfactual quantiles and, more generally, the entire distribution of the counterfactual outcome for the treated group in non-experimental settings with multiple groups and periods, where treatment timing varies. Many non-experimental designs in empirical research deviate from the canonical $\left(2\times2\right)$ scenario. A strand of the literature on causal inference has recently focused on how departures from the $\left(2\times2\right)$ scenario influence the estimation of the ATT \citep{Borusyak2021, Callaway2021, Sun2021, Wooldridge2021, DeChaisemartin2022}.\footnote{See, for instance, \cite{Roth2022} for a review.}  

While several papers provide methods for obtaining a consistent estimator of the ATT in staggered policy rollout contexts, little is known about how to estimate the distributional effect in these settings. The only study I am aware of that addresses the identification of the QTT in staggered DiD settings is \cite{LiLin2024}. Although their identification result is similar to the one presented in this paper, they do not offer an estimator for the QTT, limiting the practical applicability of their findings. A detailed comparison with their work is provided at the end of this section, following the presentation of the identification results in this paper.

To identify and estimate the full counterfactual distribution in settings with staggered treatment adoption, I exploit the intuition behind the estimator of the group-time average treatment effects proposed by \cite{Callaway2021}. By applying to each pair of treated cohorts and never-treated units the method proposed by \cite{CallawayLi2019}, I show that identification and estimation of the entire counterfactual distribution can be achieved. 

Specifically, \cite{CallawayLi2019} achieve identification and estimation of the counterfactual distribution under a distributional version of the PT assumption. The authors require the change in the untreated potential outcome to be independent of treatment assignment. The similarity of the distributional PT assumption with the canonical (mean) PT assumption makes this method very intuitive. 

Compared to the selection on observables and strong ignorability assumptions invoked by \cite{Firpo2007}, the distributional parallel trends allow unobservable confounders to vary between treated and untreated units. For instance, in the context analyzed by \cite{BonhommeSauder} of how selective and non-selective secondary education impact children's test scores, selection on unobservables means that a child's unobserved initial endowment can be (potentially) correlated with the type of education in which the pupil enrolls. 

Despite the distributional PT assumption being more than sufficient to reach point identification of the ATT, \cite{Fan2012} show that the distributional treatment effect on the treated is only partially identified. Partial identification arises from the fact that the dependence (or copula) between the change in untreated potential outcome and the pre-treatment level of untreated potential outcome is unknown to the researcher. To reach point identification, \cite{CallawayLi2019} impose a copula stability assumption, which requires this (missing) dependence to be stable over time. Under the availability of (at least) two pre-treatment periods, the researcher can recover the missing dependence using the (known) dependence in the previous periods. 

One drawback of this approach, however, is that it requires access to panel data and at least two pre-treatment periods. To overcome this requirement, following \cite{Callawayetal2018}, I impose a \textit{copula invariance} assumption, which implies that the missing dependence structure is the same for treated and never-treated units.

For instance, suppose that among never-treated units, the largest earnings increases over time occur for those in the lower deciles of the pre-treatment income distribution. The copula invariance assumption requires that, in the absence of the policy, the same pattern would have emerged for treated units.

Compared to assuming copula stability, adopting copula invariance has the additional advantage of extending the analysis to settings where repeated cross-sections, rather than panel data, are available. Under the additional assumption of rank invariance of potential outcomes over time, I demonstrate that the main identification results of this paper can indeed be extended to cases where only repeated cross-sections are available.

It is important to emphasize that the copula invariance assumption does not restrict the marginal distributions; it solely requires that the dependence structure remain the same across treated and never-treated units.

Furthermore, similarly to \cite{CallawayLi2019}, I extend the results obtained to the case where the distributional PT and the copula invariance assumption may hold after conditioning on observable characteristics.

Once the entire counterfactual distribution is identified, the researcher can construct different causal parameters of interest. For instance, a generalization of the QTT estimator proposed by \cite{CallawayLi2019} to the multi-periods and multiple groups set-up can be constructed. I call this parameter \textit{cohort-time quantile treatment effect} in the spirit of the \cite{Callaway2021}'s causal parameters. 

When computing the QTT, however, the researcher implicitly invokes an additional assumption: rank invariance between treated and untreated groups. As all these studies exploit information observed in the untreated group to infer the counterfactual distribution for the treated group, the rank invariance assumption states that treated and untreated units occupying the same ranking are comparable. However, the fact that treated and untreated units have the same ranking does not necessarily mean that rank is preserved when, for instance, treated units are endowed with untreated units' abilities. As \cite{Maasoumi2019} show, the data often reject rank invariance and, therefore, is hard to justify in most empirical applications.

For this reason, in this paper, building on \cite{Maasoumi2019}, I demonstrate that once the entire counterfactual distribution is estimated, researchers can construct causal estimands based on inequality measures -- such as the Lorenz curve, Gini coefficient, and others. For instance, one can define a causal estimand by contrasting the Gini indices computed from the cumulative distribution functions (CDFs) of the treated and untreated potential outcomes for the treated group. Inequality measures characterize the quantiles of the distribution of interest \textit{anonymously}--that is, without specifying the identity of units occupying a particular quantile. This approach addresses the limitation of rank invariance by focusing on the distributions of treated and untreated potential outcomes rather than on specific units.

However, since there is no universally accepted distribution evaluation function for conciseness, this paper will focus solely on \textit{tests for stochastic dominance rankings} to assess the distributional treatment effect on the treated. To the best of my knowledge, this is the first study to integrate the literature on causal inference with that on inequality measures. 

Depending on the causal parameter of interest, I advocate for the most appropriate statistical test to conduct valid inference. For instance, when considering the QTT, the empirical bootstrap proposed by \cite{CallawayLi2019} will be considered.

Once the causal parameters of interest are built, one can aggregate these parameters using aggregation schemes identical to those proposed by \cite{Callaway2021} to highlight heterogeneity along specific dimensions (such as how the treatment effects vary with the length of exposure to the treatment). This step is particularly crucial when there are many parameters to estimate, and the researcher wants to have a tool to aggregate and summarize results to highlight heterogeneity along a given dimension (e.g., over time) and/or to assess the overall impact of the policy.

Based on the identification results of this paper, estimating the entire counterfactual distribution is simple as it relies on non-parametric estimation of empirical distribution functions. When the distributional PT is likely to hold after conditioning on pre-treatment characteristics, estimation of the counterfactual distribution is performed via a generalization of the Inverse Probability Weighting estimator first proposed by \cite{Firpo2007} and then readapted by \cite{CallawayLi2019}. When the copula invariance assumption also holds conditionally, one can exploit the intuition behind \cite{Callawayetal2018}. Specifically, assuming that covariates are discrete, one can compute the conditional counterfactual distribution for each possible value of the covariates. 

Via different Monte Carlo exercises, I show that, despite being slightly biased for relatively small sample sizes, the proposed method's performance increases substantially as the sample size increases. In addition, I show that the performance of this method is robust to small deviations from the main identifying assumptions. 

Given the similarity between the identification results in this paper and those in the concurrent study by \cite{LiLin2024}, it is important to highlight the key differences. First, while I generalize \cite{CallawayLi2019}, they extend the method proposed in \cite{Callawayetal2018}. However, both approaches rely on a Distributional PT assumption and a Copula Invariance assumption for identification. The use of Copula Invariance in both papers enables extending the analysis to settings with repeated cross-sections.

A key distinction is that \cite{LiLin2024} derive identification results using a not-yet-treated comparison group, whereas my analysis focuses on a never-treated group. Although my main identification results are based on a never-treated comparison group, they naturally extend to settings where not-yet-treated units are used instead, due to the symmetry of the underlying arguments. Using not-yet-treated units expands the pool of valid comparisons and can improve inference by leveraging units likely to share similar pre-treatment trends with already-treated groups. However, this approach has important limitations. If later-treated units begin to adjust their behavior following the policy's introduction, the never-treated group may provide a more reliable counterfactual.

Additionally, while both studies rely on Copula Invariance for identification, I explicitly acknowledge the computational challenges when requiring the assumption to hold conditionally. As noted in \cite{CallawayLi2019}, estimating the QTT under Conditional Copula Invariance requires estimating five conditional distributions, which is often infeasible in empirical applications. For this reason, following \cite{CallawayLi2019}, I provide three different identification results depending on whether: (i) both the Distributional PT and Copula Invariance assumptions hold unconditionally, (ii) the Distributional PT assumption holds conditionally on covariates while the unconditional Copula Invariance is true, or (iii) both assumptions hold conditionally on observed covariates. In the third scenario, I demonstrate how to address the computational challenge mentioned above.

While \cite{Callawayetal2018} allow the Copula Invariance to hold conditionally on covariates, their identification results pertain to the conditional QTT. In contrast, \cite{LiLin2024} focus on the unconditional QTT. Importantly, implementing the method of \cite{Callawayetal2018} requires covariates to be discrete. Since \cite{LiLin2024} neither propose an estimator for the QTT, as I do, nor specify the nature of the covariates, it remains unclear how one might estimate the unconditional QTT under Conditional Copula Invariance. This limits the practical applicability of their identification results.

Finally, compared to \cite{LiLin2024}, I propose aggregation schemes--similar to those in \cite{Callaway2021}--to capture heterogeneity across specific dimensions. I also emphasize that once the entire counterfactual distribution is identified, researchers can derive a range of parameters beyond the QTT, which typically relies on the strong assumption of rank invariance.

Thus, while the identification results in this paper are similar to those in \cite{LiLin2024}, this paper offers greater generality and broader applicability.

Because I focus on non-parametric estimation of distributional treatment effect in a context with multiple periods and multiple groups, this paper is linked to different strands of literature. First, as pointed out in \cite{CallawayLi2019}, it contributes to the literature on nonseparable panel data methods \citep[see, for instance,][]{Chernozhukov2013}. Second, it is related to two strands of the literature in causal inference: the one which focuses on the estimation of distributional treatment effects while still allowing for selection on unobservables \citep{AtheyImb2006, BonhommeSauder, Callawayetal2018, CallawayLi2019, Miller2023}; as well as the one on staggered treatment adoption \citep{Borusyak2021, Callaway2021, Sun2021, Wooldridge2021, DeChaisemartin2022}. Lastly, since I also consider parameters that do not require rank invariance assumption to hold (such as tests of stochastic dominance), this paper contributes to the ongoing debate on inequality measures \citep[see, for instance,][]{Maasoumi2019}. 

This paper is organized as follows. Section \hyperref[Identification]{\ref*{Identification}} presents the main identification results and introduces different aggregation schemes to capture heterogeneity along specific dimensions. Section \hyperref[Estimation_2]{\ref*{Estimation_2}} outlines the estimation and inference procedures. Section \hyperref[Simulation]{\ref*{Simulation}} evaluates the finite-sample properties of the proposed estimators through various Monte Carlo exercises. In Section \hyperref[Discussion_2]{\ref*{Discussion_2}}, I compare the proposed estimator with existing methods in the literature on distributional treatment effects, and the estimator proposed in \cite{Callaway2021}. Finally, Section \hyperref[Conclusion_2]{\ref*{Conclusion_2}} provides concluding remarks.

All proofs are presented in Appendix A, while Appendix B extends the analysis to settings with repeated cross-sections.

%% file: Identification.tex
\section{Identification}\label{Identification}
This section presents the main identification results of this paper. Specifically, I will provide identification of the full distribution of the untreated potential outcome for the treated group, both when the identifying assumptions hold unconditionally and when they hold after conditioning on observed characteristics.

Throughout the paper, for any two random variables $T$ and $Z$, I denote the support of $T$ by $supp(T)$, and the support of $T$ given $Z$ by $supp(T|Z)$. 

\subsection{Setup}

Let us consider the case where there are available $T$ periods, and let us denote a generic period with $t=1,\dots, T$. Let $D_{i,t}$ denote a binary treatment indicator that equals $1$ if unit $i$ is treated in period $t$ (where $i =1,\dots, N$), and $0$ otherwise. In the classic $\left(2\times2\right)$ DiD scenario, $T=2$ and no units are treated in $t=1$. However, in the context analyzed in the rest of this paper, I will allow $T>2$. In particular, I will assume that starting from period $q \geq 2$, there is a staggered policy rollout, where $q$ denotes the first period the policy intervention is implemented. Then, one should make the following assumption about the treatment process:

\begin{assumption}[\textit{Irreversibility of Treatment}]\label{as:irreversibility}
 $D_j=0$ for all $j=1,\dots,(q-1)$ almost surely (a.s.). For $t=q,\dots,T$, $D_{t-1}=1$ implies that $D_t=1$ a.s.
\end{assumption}

\hyperref[as:irreversibility]{Assumption \ref*{as:irreversibility}} states that no unit is treated before period $q$ and that once a unit receives the treatment, it will remain treated for the remainder of the panel. This assumption is often referred to as \textit{staggered treatment adoption}. Following \cite{Callaway2021} and \cite{Sun2021}, I will interpret this assumption as units changing their behavior "forever" once they become treated.\footnote{If always-treated units exist, these will be dropped from the analysis, as there is no pre-treatment period for these units.}$^{,}$\footnote{For an analysis of the case where the treatment is binary, the parameter of interest is the ATT, and the treatment switches on and off at different points in time, see \cite{DeChaisemartin2022}.}

Another implication of \hyperref[as:irreversibility]{Assumption \ref*{as:irreversibility}} is that it automatically defines the "cohort" to which unit $i$ belongs. Assuming a never-treated group exists, one can define $T-q+2$ mutually exclusive cohort dummies, $d_r$ (with $r=q,\dots, T$), denoting the period in which unit $i$ first receives the treatment. On the other hand, $C$ will be a dummy taking value $1$ if a group is never-treated, and $0$ otherwise. Lastly, $\Bar{d}=\text{max}_{i=1,\dots, N} d_{i,r}$ denotes the maximum period in the dataset where a unit may become treated. 

For brevity, I will present the main identification results focusing on the setting with a never-treated comparison group. These results can be generalized to cases where not-yet-treated units serve as the comparison group, as they follow from symmetric arguments  \citep{Callaway2021}. While using not-yet-treated units expands the pool of valid comparison units--potentially improving inference due to their likely similarity in pre-treatment trends with already-treated units--this approach has an important drawback. If units that are treated later adjust their behavior in response to the policy's introduction, the never-treated group may provide a more reliable counterfactual.\footnote{See \cite{Ciaccio2023} for an empirical application where never-treated units served as a more appropriate comparison group, as some later-treated units altered their behavior to delay the policy's implementation. As discussed in the next section, such behavior constitutes a violation of the Limited Treatment Anticipation assumption, which is necessary for identifying the estimands considered in this paper.}

Next, assume the existence of a complete set of pre-treatment covariates, denoted by $X$ with support denoted by $\chi = supp(X) \subseteq  R^K$ and dimensionality equal to $dim(X)=k$. The generalized propensity score can then be defined as $\P_{r,t}(X)=\P\left(d_r=1|X, d_r+C=1\right)$, which represents the probability of being first treated in period $r$, conditional on observable characteristics and on either belonging to cohort $r$ or being never-treated.\footnote{When using not-yet-treated units as the comparison group, the generalized propensity score takes the form $\P_{r,s}(X)=\P\left(d_r=1|X, d_r+(1-D_s)(1-d_r)=1 \right)$. Here, $\P_{r,s}(X)$ denotes the probability of being first treated in period $r$ , conditional on observable characteristics and on either belonging to cohort $r$ (i.e., $d_r=1$) or being in the not-yet-treated group by time $s$ (i.e., $(1-D_s)(1-d_r)=1$).}

We can set up a potential outcomes framework. Specifically, as in \cite{Callaway2021}, I will combine the dynamic potential outcomes framework \citep{Robins1986, Robins1987} with the dynamic treatment adoption setting \citep{Heckmanetal2016}. For $r=q, \ldots, T$, let $ Y_{i,t}(r)$ represent the potential outcome for unit $i$ at time $t$ had the policy been introduced by period $r$ (i.e., $d_{i,r}=1$). Whereas $ Y_{i,t}(0)$ will denote the analogous in period $t$ had the treatment not been received (that is, had the unit never been treated). Then, one can rewrite the observed outcome for a generic unit as follows: 
\begin{equation}
     Y_{i,t} =  Y_{i,t}(0)+ \sum^T_{r=q} d_{i, r} \cdot \left( Y_{i,t}(r)-  Y_{i,t}(0)\right) 
\end{equation}

That is, for each unit, we observe only one of the two mutually exclusive potential outcomes, depending on whether the unit receives the treatment.  Specifically, the observed outcome corresponds to the untreated potential outcome, $Y(0)$, for units that are never treated  (for all $t=1,\dots, T$). For units first treated in period $r$, the observed outcome equals $Y(0)$ for all $t<r$, and equals $Y(r)$ for all $t\geq r$. The fact that we can only observe one of the two potential outcomes--either $Y(r)$ or $Y(0)$--for any given unit at any given time creates a selection problem. This issue is commonly referred to as \textit{fundamental problem of causal inference} \citep{Holland1986}.

I will also make the following assumption for the remainder of the paper: 
\begin{assumption}[\textit{Random Sampling}]\label{as:RS}
 $\{Y_{i,1}, Y_{i,2},..., Y_{i,t}, X_i, D_{i,1}, D_{i,2},...,D_{i,\tau}\}^n_{i=1}$ is independent and identically distributed.
\end{assumption}

This assumption requires the availability of panel data. In Appendix B, I extend the analysis to settings where repeated cross-sections, rather than panel data, are available. To do so, I follow \cite{Callawayetal2018} and \cite{Callaway2021}.\footnote{For an analysis when balanced panel data is not available, see \cite{bellegoetal2025}.}

\hyperref[as:RS]{Assumption \ref*{as:RS}} is a common assumption in the Diff-in-Diff literature. It allows us to introduce uncertainty in our setting and consider potential outcomes as random variables. Specifically, \hyperref[as:RS]{Assumption \ref*{as:RS}} implies that any unit $i$ is the realization from a super-population of interest. It is worth stressing that, however, \hyperref[as:RS]{Assumption \ref*{as:RS}} neither rules out any time series dependence nor imposes any restriction between the relation between $D_{i,t}$ and $(Y(0),Y(r))$.

Under Assumptions \hyperref[as:irreversibility]{\ref*{as:irreversibility}} and \hyperref[as:RS]{\ref*{as:RS}}, it is possible to define the estimand of interest. Furthermore, it is worth emphasizing that assuming random sampling rules out the possibility of interference--commonly referred to as the \textit{no interference} assumption. This means that the potential outcome of unit $i$ is not affected by the treatment status of unit $j\neq i$. 

The no interference assumption, together with the assumption of a well-defined treatment (i.e., \textit{treatment consistency}) constitute the \textit{Stable Unit Treatment Value Assumption (SUTVA)}.\footnote{For a discussion on potential violations of SUTVA and approaches to account for spillover effects, see \cite{Butts2021, Fiorini2024, xu2025}.}

Before introducing the assumptions needed for identification, I will now introduce the notation both for the quantile and the tests of stochastic dominance ranking. This will be crucial since we are interested in understanding the impact of the policy along with the distribution of our outcome of interest.  For $\tau \in \left[0,1\right]$ the $\tau$th quantile $q_{\tau}$ of a generic random variable T is defined as:
    \begin{equation*}
        \P(T\leq q_{\tau})=F_T(q_{\tau})=\tau
    \end{equation*}

or analogously as $q_{\tau}=F^{-1}_T(\tau):=inf\{t:F_T(t)\geq\tau\}$. Where, for instance, $\tau=.25$ represents the $25^{th}$ percentile.

However, summarizing the distributional treatment effects by comparing quantiles via a quantile-by-quantile approach relies implicitly on the rank invariance assumption. That is, treated and never-treated units with the same rank also have the same unobserved characteristics. As \cite{Maasoumi2019} show, empirical evidence usually dismisses this assumption. 

To address this issue, I recommend an alternative approach that first summarizes the impact by an appropriate evaluation function, and then computes the difference between the resulting evaluations--as suggested by \cite{Maasoumi2019}. This contrasts with the quantile-by-quantile approach, which involves computing individual quantiles for the treated and untreated potential outcomes and then taking the difference between them.
 
There is no unique definition of the distribution's evaluation function: "averages, inequality measures, and entropies are all well-known functions of distributions that summarize its quantiles anonymously, without regard to the identity of those who occupy a given quantile. Each function attributes its own weights to different wage levels." \citep[p. 2439]{Maasoumi2019}. Given the existence of multiple valid evaluation functions--often selected within a decision-theoretic framework--I adapt the statistical tests for stochastic dominance rankings proposed by \cite{Maasoumi2019} to the context under analysis.

Considering partial orderings between random variables offers two key advantages, in addition to avoiding reliance on the assumption of rank invariance. First, there is no compelling reason to privilege one inequality measure over another; choosing one over the other hinges upon a decision-theoretic framework. Stochastic dominance tests help overcome this limitation by identifying whether one distribution dominates another without requiring the specification of a particular weighting scheme.

Second, when the distributions of treated and untreated potential outcomes intersect, different inequality measures may yield conflicting conclusions. This issue arises even when rank invariance holds, as the sign and magnitude of the treatment effect can vary across quantiles. Consequently, the conclusions drawn depend heavily on the specific evaluation function employed. In such cases, stochastic dominance tests offer a more robust framework for policy evaluation, as they assess whether distributions can be consistently ranked across a broad range of evaluation functions with a certain level of statistical confidence.

Let us now define what tests of stochastic dominance rankings are. I will use the notation used by \cite{Maasoumi2019}. Let $U_1$ be the class of von Neumann-Morgenstern utility functions $u$ that are increasing (i.e., $u'\geq 0$) in the potential outcome of interest, $Y(d)$ with $d=\{0,1\}$. Let $U_2$ be the class of utility function in $U_1$ such that the utility function is concave ($u''\leq 0$). 

\textit{First-order Stochastic Dominance} -- The treated potential outcome, $Y_{t}(r)|d_r=1$, first-order stochastically dominates (denoted by $Y_{t}(r)|d_r=1$ FSD $Y_{t}(0)|d_r=1$) the untreated potential outcome for the treated, $Y_{t}(0)|{d_r=1}$, if and only if
\begin{enumerate}
    \item $\E u(Y_{t}(r)|d_r=1 )\geq \E u(Y_{t}(0)|d_r=1)$ for all $u \in U_1$ with the inequality being strict for some $u$, or
    \item $F_{Y_{t}(r)|d_r=1}(y) \leq F_{Y_{t}(0)|d_r=1}(y)$ for all $Y$ with the inequality being strict for some values of $Y$, or 
    \item $Y_{t,\tau}(r)|d_r=1\geq Y_{t,\tau}(0)|d_r=1$ for all points in the support of $Y$.
\end{enumerate}

where $\left(Y_{t,\tau}(r)|d_r=1\right)-\left(Y_{t,\tau}(0)|d_r=1\right)$ represents the difference in the treated and the untreated potential outcomes for the treated group at the $\tau$th quantile. 

\textit{Second-order Stochastic Dominance} -- The treated potential outcome, $  Y_{t}(r|d_r=1)$, second-order stochastically dominates (denoted by  $Y_{t}(r)|d_r=1$ SSD $Y_{t}(0)|d_r=1$) the untreated potential outcome for the treated, $Y_{t}(0)|{d_r=1}$, if and only if
\begin{enumerate}
    \item $\E u(Y_{t}(r)|d_r=1 )\geq \E u(Y_{t}(0)|d_r=1)$ for all $u \in U_2$ with the inequality being strict for some $u$, or
    \item $\int_{-\infty}^yF_{Y_{t}(r)|d_r=1}(z) dz \leq \int_{-\infty}^y F_{Y_{t}(0)|d_r=1}(z) dz$ for all $Y$ with strict inequality for some values of $Y$, or 
    \item $\int_{0}^{\tau}\left(  Y_{t}^u(r)|d_r=1\right) du \geq \int_{0}^{\tau} \left(  Y_{t}^u(0)|d_r=1\right)du$ for all points in the support of $Y$.
\end{enumerate}

where for notational convenience $(  Y_{t}^\tau(r)|d_r=1):=F^{-1}_{Y_t(r)|d_r=1}(\tau)$ represents the $\tau$th quantile of $Y_t(r)|d_r=1$.

If, for instance, $Y_{t}(r)|d_r=1$ FSD $Y_{t}(0)|d_r=1$, then it is also true that $Y_{t}(r)|d_r=1$ SSD $Y_{t}(0)|d_r=1$. This is because first-order stochastic dominance implies second-order stochastic dominance.

\subsection{The cohort-time distributional treatment effect on the treated}

As \cite{Manski2013} notice, the fact that data will reveal only one of the two potential outcomes--either $ Y_{i,t}(r)$ or $ Y_{i,t}(0)$--for any given unit at a specific time makes it challenging finding an appropriate way to recover the counterfactual outcome--the outcome that would have been observed had the policy not been implemented.

Researchers often employ various strategies to recover counterfactual outcomes and evaluate the impact of a given policy. In non-experimental settings, one of the most commonly estimated causal parameters is the ATT. Within the potential outcomes framework, the ATT is defined as $ATT=\E(Y(1)-Y(0)|D=1)$ (where $D$ is a dummy taking value $1$ is a binary indicator equal to 1 if the unit is treated). This parameter captures the average effect of the treatment on those units that actually received the treatment.

Since the counterfactual outcome for treated units is unobservable, researchers typically rely on different identifying assumptions to estimate $Y(0)$ consistently.\footnote{See \cite{Imbens2009} for a comprehensive review of these assumptions.} One of the most widely used is the Parallel Trends assumption \citep{Heckman1998}. In the canonical difference-in-differences (DiD) setup--with two groups and two time periods--the PT assumption posits that, in the absence of treatment, the average change in outcomes over time would have been the same for treated and untreated units. Formally, this can be expressed as $\E(Y(1)-Y(0)|d=1)=\E(Y(1)-Y(0)|d=0)$. This assumption allows researchers to use the untreated group's outcome evolution as a valid counterfactual for the treated group. Recent studies have extended this framework to settings with multiple time periods and staggered treatment adoption \citep{Borusyak2021, Callaway2021, Sun2021, Wooldridge2021, DeChaisemartin2022}.

However, when treatment effects are likely to be heterogeneous along the distribution of the outcome of interest, considering the ATT alone might be misleading. In such contexts, analyzing the distributional effects of the policy becomes essential for a sound policy evaluation. Researchers commonly address this by comparing the distributions of treated and untreated potential outcomes \citep[e.g.,][]{Carneiro2001}.

As discussed in \hyperref[Introduction]{Section~\ref*{Introduction}}, the distributional analog of the ATT is the Quantile Treatment Effect on the Treated. In this paper, I consider a generalization of the QTT to the setups with multiple periods and groups with variation in treatment timing. Moreover, following \cite{Maasoumi2019}, I also consider parameters based on tests of stochastic dominance rankings.

Specifically, if we denote with $F_{Y_{t}(r)|d_r=1}$ and $F_{Y_{t}(0)|d_r=1}$ the distribution of treated and untreated potential outcomes for units that first receives the treatment in period $r$, then the QTT parameter for units first treated in group $r$ evaluated at any given time period $t\geq r$ can be defined as follows:

\begin{equation}\label{qtt}
       QTT_{r,t}(\tau) = F^{-1}_{Y_{t}(r)|d_r=1}(\tau)- F^{-1}_{Y_{t}(0)|d_r=1}(\tau) \quad \tau\in\left[0,1\right], r=q,\dots,T, \text{ and } t\geq r,  
    \end{equation}

Borrowing the name from \cite{Callaway2021}, I call these parameters as \textit{cohort-time distributional treatment effects on the treated}.
 
Besides allowing heterogeneity of the treatment effect along with the distribution of $Y$, the parameter $QTT_{r,t}(\tau)$ allows additional heterogeneity across treated cohorts (note that, for example, even keeping $t$ and $\tau$ fixed, the QTT experienced by cohort $r$ may be different from that experienced by cohort $r'$) and over time. Moreover, the researcher can employ the cohort-time quantile treatment effects to learn about the overall impact of the policy by constructing an overall quantile treatment effect or to highlight heterogeneity along specific dimensions (such as, how the QTT varies with the length of exposure to the treatment). 

As far as the distributional treatment effect based on tests of stochastic dominance rankings is concerned, using the definition of stochastic dominance, one can check for each post-treatment period $t$ and each treated cohort $r$ whether $Y_{t}(r)|d_r=1$ FSD $Y_{t}(0)|d_r=1$ or $Y_{t}(r)|d_r=1$ SSD $Y_{t}(0)|d_r=1$. As pointed out in the previous subsection, comparing distributions rather than units -- besides overcoming the issue of assuming rank invariance -- allows the researcher to have a clearer picture of the overall effect of the policy along the distribution of $Y$. This becomes particularly crucial when treated and untreated potential outcomes distributions for the treated group cross, making policy evaluation based on a quantile-by-quantile approach extremely sensitive to the quantile analyzed.

\subsection{Identifying assumptions}

While the Parallel Trends assumption is sufficient to retrieve a consistent estimator of the ATT in the standard $(2\times2)$ DiD framework -- and its extensions to multi-period settings have enabled identification of ATT in more complex scenarios -- identifying distributional treatment effects poses greater challenges. A straightforward, distributional analog of the PT assumption is no longer sufficient for this purpose \citep{Fan2012}.

Since the distribution $F_{Y_{t}(r)|d_r=1}$ is directly identified from the data, point-identification of the distributional treatment effect in a staggered treatment adoption setting requires identifying the counterfactual distribution $F_{Y_{t}(0)|d_r=1}$--that is, the distribution of untreated potential outcomes for units treated in cohort $r$. To achieve this, we introduce the following assumptions:

\begin{assumption}[\textit{Limited Treatment Anticipation}]\label{as:NA}
There is a known $\rho \geq 0$ such that $\P\{Y_t(r)\leq y |X, d_r=1\}=\P\{Y_t(0)\leq y|X, d_r=1\}$ a.s. for all $r=q,\dots, T$ and $t=1,\dots,T$ such that $t<r-\rho$.
\end{assumption}

\hyperref[as:NA]{Assumption \ref*{as:NA}} is an extension from the mean to the entire distribution of the Limited Treatment Anticipation assumption commonly invoked in DiD settings with staggered treatment adoption \citep[see, for instance,][]{Callaway2021}. It states that, before the treatment occurs, the treated and untreated potential outcomes of "eventually" treated units are the same in the pre-treatment period. In other words, this implies that units either do not anticipate the treatment (i.e., $\rho=0$), or if anticipation is possible, it occurs within a known and limited time window (i.e., $\rho>0$, such that the assumption holds for all $t<r-\rho$). Specifically, when $\rho = 0$, the condition reduces to the No Anticipation assumption.

It is worth stressing that \hyperref[as:NA]{Assumption \ref*{as:NA}} implies that the distributional treatment effect is $0$ for all $t<r-\rho$. Further, \hyperref[as:NA]{Assumption \ref*{as:NA}} is likely to hold in empirical practice when units do not expect the implementation of the policy and/or units cannot "decide" whether to be treated or not. 

Next, the following assumption provides a distributional version of the classic Parallel Trends assumption in staggered treatment adoption settings.

\begin{assumption}[\textit{Conditional Distributional PT based on a "Never-treated" Group}]\label{as:CPTnever}Let $\rho$ be defined as in \hyperref[as:NA]{Assumption \ref*{as:NA}}. For each $r,t \in \{q,\dots,T\}$ such that $t \geq q-\rho$, \\
        \begin{equation*}
           \P\left( \Delta Y_t(0)\leq \Delta Y|X, d_r=1\right)=\P\left(\Delta Y_t(0)\leq \Delta Y|X, C=1\right) \quad \text{a.s.}
        \end{equation*} 
\end{assumption}

where $\Delta Y_t=  Y_{t}- Y_{t-1}$. \hyperref[as:CPTnever]{Assumption \ref*{as:CPTnever}} states that, in the absence of treatment, the counterfactual distribution of potential outcomes for units treated in cohort $r$ would have evolved in parallel to that of the never-treated units over time. The \hyperref[as:CPTnever]{Distributional PT Assumption} is commonly used in the existing literature to retrieve the counterfactual distribution \citep{Fan2012, CallawayLi2019, Miller2023}, extending the idea of the common trends assumption from the average to the entire distribution of untreated potential outcomes. 

To give the intuition, suppose we want to study the effect of a state-level minimum wage increase on workers' wages. The Distributional PT Assumption implies that, by comparing states that eventually raise the minimum wage with those that never do, the path of earnings, in the absence of the policy, would have been similar between treated and never-treated states. If, for instance, the path of earnings in the absence of the policy depends on education or gender, then the Distributional PT assumption holds only after controlling for these variables. Failing to control for these covariates would lead to biased estimates. 

On the other hand, this assumption may fail if different parts of the outcome distribution evolve differently across groups before treatment (e.g., the lower tail of the income distribution in the treated group grows faster than in the control group). For instance, before the minimum wage increase, low-wage workers in treated states may have already experienced faster wage growth due to local labor shortages, while in never-treated states, wages remained relatively stagnant. In this case, even without the minimum wage increase, the wage distribution in the treated state would have evolved differently from that in the control states. Assuming distributional parallel trends in this scenario could lead to incorrect attribution of pre-existing wage growth or labor market changes to the minimum wage policy, resulting in biased estimates of the treatment effect.\footnote{If the assumption fails, bias will vary across quantiles, leading to overestimated effects at some quantiles and underestimated effects at others.}

It is important to note that while the common mean PT assumption requires that trends be similar on average between treated and never-treated cohorts, the distributional PT assumption requires that the entire distribution evolves similarly over time.

Unlike selection-on-observables invoked by \cite{Firpo2007},\footnote{The selection-on-observable assumption--also known in the program evaluation literature as unconfoundedness assumption--requires $Y(0)$ and $Y(1)$ to be independent of treatment assignment, once controlling for a complete set of pre-treatment characteristics.} the distributional parallel trends also allow unobservable confounders to vary between treated and untreated cohorts. Allowing for selection on unobservables makes \hyperref[as:CPTnever]{Assumption \ref*{as:CPTnever}} weaker. It is also weaker than Assumptions 1 and 2 in \cite{BonhommeSauder}, which require that the unobserved time-varying shock in the pre-treatment period and the time-varying unobserved component in the equation for the untreated potential outcome in the post-treatment period be independent of treatment assignment and independent of unobservable confounders (once controlling for $D_i$). 

As \cite{Callawayetal2018} point out, if the pre-treatment period is long, then \hyperref[as:CPTnever]{Assumption \ref*{as:CPTnever}} is testable under the assumption of strict stationarity of the time series of changes in untreated potential outcomes. 

Despite being stronger than the mean PT assumption--and more than sufficient to achieve point identification of the ATT--\footnote{Once the counterfactual distribution is estimated, the average counterfactual distribution for the treated cohort can be computed as $\E(Y(0)|d_r=1)=\int^1_0 F^{-1}_{Y(0)|d_r=1}(\tau)d\tau$. The ATT can then be straightforwardly retrieved.}\hyperref[as:CPTnever]{Assumption \ref*{as:CPTnever}} is no longer sufficient to achieve point identification of $F_{Y_{t}(0)|d_r=1}$. Exploiting results on the distribution of the sum of two random variables, \cite{Fan2012} show that the Distributional PT assumption is insufficient to point-identify $F_{Y_{t}(0)|d_r=1}$, making it more challenging to identify the distributional treatment effect. Without additional assumptions, $F_{Y(0)|d_r=1}$ can only be partially identified \citep{CallawayLi2019}.

The reason why \hyperref[as:CPTnever]{Assumption \ref*{as:CPTnever}} is insufficient for point identification of the distributional treatment effect is that different distributions of untreated potential outcomes in period $t$ may be observationally equivalent. As \cite{Callawayetal2018} highlight, an additional assumption on the dependence structure (or copula) between $Y_{t-1}(0)$ and $\Delta Y_t(0)$ is required. Even if the marginal distributions $F_{Y_{t-1}(0)|d_r=1}$ and $F_{\Delta Y_t(0)|d_r=1}$ can be identified separately, since "observations are observed separately for untreated and treated individuals,\dots, the joint distribution is not identified" \cite[p.1585]{CallawayLi2019}. 

To retrieve the joint distribution of $Y_{t-1}(0)$ and $\Delta Y_{t}(0)$, following \cite{CallawayLi2019}, I will employ Sklar's theorem \citep{Sklar1959} which states that any joint distribution can be expressed as a function (or \textit{copula}) of its marginal distributions. Specifically, for any two random variables $T$ and $W$, Sklar's theorem says that the joint distribution, $F_{T,W}$, can be written as:
\begin{equation*}
    F_{T,W}= C_{T,W}(F_T(t),F_W(w))
\end{equation*}

where $C_{T,W}(\cdot,\cdot)$ is the copula.\footnote{The copula captures all the dependence structure between the two random variables but contains no information about their marginal distributions. For a detailed discussion on copulas, see, for example, \cite{Nelsen2006}.}. 

Copulas are widely used in nonparametric statistics because they allow researchers to model dependence between two random variables independently of their marginal distributions. Moreover, "much of the usefulness of copulas in the study of nonparametric statistics derives from the fact that for strictly monotone transformations of the random variables, copulas are either invariant or change in predictable ways" \citep[p. 25]{Nelsen2006}.

The final step is to specify the copula function in this setting. In particular, following \cite{Callawayetal2018}, I will assume that the dependence structure is the same for treated units in cohort $r$ and never-treated units:  

\begin{assumption}[\textit{Conditional Copula Invariance based on a "Never-treated" Group}]\label{as:Copula} For all $x \in \chi$ and for all $(u,v)\in \left[0,1\right]^2$
        \begin{equation*}
            C_{\Delta Y_{t}(0), Y_{t-1}(0)| X, d_r=1}\left(u,v\right)= C_{\Delta Y_{t}(0), Y_{t-1}(0)| X, C=1}\left(u,v\right)
        \end{equation*}
\end{assumption}

\hyperref[as:Copula]{Assumption \ref*{as:Copula}} differs from the assumption in \cite{CallawayLi2019}, who instead impose that the copula between $Y_{t-1}(0)$ and $\Delta Y_{t}(0)$ remains stable over time. The key advantage of adopting Conditional Copula Invariance over the Conditional Copula Stability assumption in \cite{CallawayLi2019} is that it does not require access to panel data with at least two pre-treatment periods. This flexibility allows researchers to work with repeated cross-sections, as I demonstrate in Appendix B. 

Specifically, \hyperref[as:Copula]{Assumption \ref*{as:Copula}} recovers the missing dependence between $Y_{t-1}(0)$ and $\Delta Y_{t}(0)$ by replacing the unknown copula, $C_{\Delta Y_{t}(0), Y_{t-1}(0)| X, d_r=1}\left(\cdot,\cdot\right)$, with the "observed" dependence from the never-treated group. 

Intuitively, in the example given above about the increase in the state-level minimum wage, \hyperref[as:Copula]{Assumption \ref*{as:Copula}} implies that the dependence between the change in wages over time and the pre-treatment level of wages is identical between treated and never-treated units in the absence of the policy. If, for example, the dependence between the change in wages over time and the initial level of wages is influenced by the level of education, then for \hyperref[as:Copula]{Assumption \ref*{as:Copula}} to hold, we should condition on educational attainment.

Let us now consider a case where \hyperref[as:Copula]{Assumption \ref*{as:Copula}} is likely to be violated. Consider again the increase in the state-level minimum wage, and suppose we are interested in evaluating its impact on wages for low-wage workers. Suppose that, by time $t$, all states have maintained a stable minimum wage for over a decade. For simplicity (though this is not necessary), assume that all states face the same wage level for low-wage individuals, $w$. Now, suppose that minimum wage is first increased by state $s$ in period $t+1$. However, in the year prior to this increase, between periods $t-1$ and $t$, wages for low-wage workers rise much more in the states that, during the observed period, do not raise the minimum wage (i.e., never-treated states) due to local labor shortages. In contrast, in the eventually later-treated states, wages for low-wage workers remain almost stagnant. 

In this context, the copula represents the dependence between the initial wage level $w$ and the subsequent wage increase occurring between periods $t-1$ and $t$. Since this dependence differs between treated and never-treated states even before the minimum wage increase, \hyperref[as:Copula]{Assumption \ref*{as:Copula}} is likely to be violated. Using this method will, therefore, lead to biased estimates. The extent of the bias will depend on the difference in the dependence described above between never-treated and eventually treated units. The larger the difference in dependence, the greater the bias.

It is important to highlight that \hyperref[as:Copula]{Assumption \ref*{as:Copula}} imposes no restrictions on the marginal distributions; it only restricts the dependence (i.e., the initial distributions can indeed differ between treated and never-treated groups).

Moreover, as noted by \cite{Callawayetal2018} and \cite{CallawayLi2019},  \hyperref[as:CPTnever]{Assumption \ref*{as:CPTnever}} does not imply \hyperref[as:Copula]{Assumption \ref*{as:Copula}}, nor the reverse is true. The reason the two assumptions are not mutually implied is that while \hyperref[as:CPTnever]{Assumption \ref*{as:CPTnever}} restricts only the marginals of the change in untreated potential outcomes,  \hyperref[as:Copula]{Assumption \ref*{as:Copula}} restricts the dependence between $Y_{t-1}(0)$ and $\Delta Y_{t}(0)$.  To illustrate this, suppose that $Y_{t-1}(0)$ and $\Delta Y_{t}(0)$ are jointly normally distributed with means $\mu_{1}$ and $\mu_{2}$ respectively, and variances $\sigma^2_1$ and $\sigma^2_2$, respectively. If the correlation between $Y_{t-1}(0)$ and $\Delta Y_{t}(0)$ is denoted as $\rho_t$, then under \hyperref[as:Copula]{Assumption \ref*{as:Copula}} we require that $\rho_t$ be independent of treatment assignment, conditionally on $X$ and either being treated in cohort $r$ or never-treated. This assumption does not restrict the means or variances of the two random variables, but only their correlation. 

While no formal tests have been proposed thus far to check whether \hyperref[as:Copula]{Assumption \ref*{as:Copula}} hold, \cite{CallawayLi2019} suggests using rank correlation measures, such as Kendall's $\tau$, to assess its plausibility in the pre-treatment periods. Using a test based on Kendall's $\tau$, they show the validity of their Copula Stability assumption when evaluating the effect of increasing minimum wages on the unemployment rate in the U.S. during years 2000-2007, thus providing evidence of contexts where assumptions like \hyperref[as:Copula]{Assumption \ref*{as:Copula}} are likely to hold.

On the one hand, the researcher can inspect whether the dependence between the initial level of outcomes and the change in outcomes over time is similar between treated and never-treated units. However, if such dependence is more likely to be similar among specific groups within the population, the researcher should condition on these characteristics. Conditioning on observable pre-treatment characteristics should increase the likelihood that \hyperref[as:Copula]{Assumption \ref*{as:Copula}} holds in practice.

Under \hyperref[as:CPTnever]{Assumption \ref*{as:CPTnever}} and \hyperref[as:Copula]{Assumption \ref*{as:Copula}} we can point-identify $F_{Y(0)|d_r=1}$. However, since the copula representation is not unique unless the random variables are continuous \citep{joe1997multivariate, Nelsen2006}, following \cite{Callawayetal2018} and \cite{CallawayLi2019}, I will impose the following additional assumption:

\begin{assumption}[\textit{Continuity}]\label{as:Continuity} The random variables $Y_{t-1}(0)$ and $\Delta Y_t(0)$ have continuous distribution conditional on either being part of the treated cohort $r$ (i.e., $d_r=1$) or being never-treated (i.e., $C=1$), and $Y_{t}(r)|d_r=1$ is also continuously distributed on its support. Moreover, each of these distributions has a compact support with marginal distributions, which are (uniformly) bounded away from $0$ and $1$ over their respective support.
\end{assumption}

Lastly, the following assumption will be required only to recover $F_{Y(0)|d_r=1}$ when conditioning on observable pre-treatment covariates:

\begin{assumption}[\textit{Overlap}]\label{as:Overlap} For each $r,t \in \{q,...,T\}$, there exists some $\varepsilon>0$ s.t. $\P(d_r=1)>\varepsilon$ and $\P_{r,t}(X)<1-\varepsilon$ a.s.
\end{assumption}

where $\P(d_r=1)$ denotes the probability of being treated in cohort $r$, and $\P_{r,t}(X)=\P\left(d_r=1|X, d_r+C=1\right)$ denotes the generalized propensity score. Specifically, while $\P(d_r=1)$ captures the fraction of treated units in a given cohort, $\P_{r,t}(X)$ represents the probability of belonging to cohort $r$ as a function of observable characteristics, and either being part of cohort $r$ or the never-treated group.

\hyperref[as:Overlap]{Assumption \ref*{as:Overlap}} is the common overlap condition and generalizes the assumptions done in \cite{Firpo2007, Callawayetal2018, CallawayLi2019} to the multiple periods and multiple groups setting. In words, the first part of \hyperref[as:Overlap]{Assumption \ref*{as:Overlap}} states that there is a positive probability of being treated in cohort $r$ (compared to being always never-treated). The second part, instead, says that, for all the combinations of $(r,t)$, the propensity score is bounded away from $1$.


\subsection{Non-parametric identification of the distribution of untreated potential outcome}
In this subsection, I will show that, under the aforementioned assumptions, the full counterfactual distribution for the treated group, $F_{Y(0)|d_r=1}$, can be non-parametrically point-identified in Difference-in-Differences settings with multiple periods and multiple groups. 

Suppose absurdly that pre-treatment covariates play no role in the identification. The following theorem shows that under Assumptions \hyperref[as:irreversibility]{ \ref*{as:irreversibility}}, \hyperref[as:RS]{\ref*{as:RS}}, \hyperref[as:Continuity]{\ref*{as:Continuity}} and an unconditional version of Assumptions \hyperref[as:NA]{\ref*{as:NA}}--\hyperref[as:Copula]{\ref*{as:Copula}}, we can point identify $F_{Y(0)|d_r=1}$.

\begin{theorem}\label{Th1}
\textit{Suppose Assumptions 1, 2, 6, and unconditional version of Assumptions 3-5 hold. Then} $F_{Y_{t}(0)|d_r=1}\left(\tau\right)$ \textit{is identified}:
        \begin{equation*}
         \begin{aligned}
             F_{  Y_{t}\left( 0 \right)|d_r=1}  & = \P\left(Y_t \left(0\right) \leq y| d_r=1\right) \\
            & = \E \left[ \mathbbm{1} \left(\Delta_{\left[r-\rho-1,t\right]}Y(0)\right)\right. \leq y - \\
            & \left. \left. F^{-1}_{Y_{r-\rho-1}\left(0 \right) |d_{r}=1} \left(F_{Y_{r-\rho-1}\left(0\right) |C=1} \left(Y_{r-\rho-1}(0)\right)\right)\right]  |C=1 \right]
         \end{aligned} 
         \end{equation*}
\end{theorem}

where $\Delta_{\left[r-\rho-1,t\right]}Y(0)=Y_t \left(0\right)-Y_{r-\rho-1} \left(0\right)$ represents the long difference. 

\hyperref[Th1]{Theorem \ref*{Th1}} is the main result of this paper. It shows that, under the classical assumptions done in the staggered Diff-in-Diff literature, an extension of the common PT assumption to the entire distribution of change in untreated potential outcomes and a "new" assumption regarding the joint distribution of $Y_{t-1}(0)$ and $\Delta Y_t(0)$, one can point reach point-identification of $F_{Y_{t}(0)|d_r=1}\left(\tau\right)$. Once $F_{Y_{t}(0)|d_r=1}\left(\tau\right)$ is identified, we can also identify $F^{-1}_{Y_{t}(0)|d_r=1}\left(\tau\right)$. 

\hyperref[Th1]{Theorem \ref*{Th1}} implies that units belonging to treated cohort $r$ must be similarly distributed to never-treated units in terms of both marginals distributions of $Y(0)_{t-1}$ and $\Delta Y(0)_{t}$, but also in terms of the dependence between these marginals. This is guaranteed by Assumptions 4 and 5.\footnote{It is worth stressing that, under Assumptions 1-4, it is also possible to identify the ATT in a context with multiple periods and groups with staggered treatment adoption.}

The intuition underlying the result in \hyperref[Th1]{Theorem~\ref*{Th1}} closely follows that of \cite{CallawayLi2019} for their main identification results. Specifically, note that:

$$\P\left(Y_t \left(0\right) \leq y| d_r=1\right)= \E \left[ \mathbbm{1} \left(\Delta_{\left[r-\rho-1,t\right]}Y(0) + Y_{r-\rho-1} \left(0\right)\right) \leq y|d_r=1\right]$$

is an integral (remember that $Y$ is continuous) over the joint distribution between the pre-treatment level in the untreated potential outcome, $Y_{r-\rho-1}$, and the change in untreated potential outcome. This joint distribution can be identified under \hyperref[as:Copula]{Assumption \ref*{as:Copula}}, which allows recovering the missing dependence by replacing the unknown copula for the treated with that observed for the never-treated. However, since $\Delta_{\left[r-\rho-1,t\right]}Y(0)$ is not observed for the treated group $r$, we need \hyperref[as:CPTnever]{Assumption \ref*{as:CPTnever}} to replace $F^{-1}_{\Delta_{\left[r-\rho-1,t\right]}Y(0)|d_r=1}(\cdot)$ with $F^{-1}_{\Delta_{\left[r-\rho-1,t\right]}Y(0)|C=1}(\cdot)$. For the entire proof, please refer to Appendix A.

The following example shows the conditions for a classic two-way fixed effects regression to satisfy the assumptions needed to identify the group-time average treatment effects proposed by \cite{Callaway2021} and the distribution of the untreated potential outcome presented in this paper.

\texttt{Example 1.} Consider the following two-way fixed effects regression for the untreated potential outcome:
\begin{equation*}
     Y_{it}(0)= \alpha_t + \eta_i + u_{it} 
\end{equation*}

where $\alpha_t$ represents the time-fixed effects, $\eta_i$ is the unobserved heterogeneity -- which may be distributed differently between treated cohorts and never-treated units --, and $u_{it}$ represents the time-varying unobservable shock. Suppose we have access to panel data and assume, for simplicity, there is no treatment anticipation (i.e., $\rho=0$). For this data-generating process (DGP) to satisfy the assumptions of the estimator of the $ATT$ presented in \cite{Callaway2021}, a sufficient condition for the mean version of \hyperref[as:CPTnever]{Assumption \ref*{as:CPTnever}} to hold would be that $\E(\Delta u_{it}|d_r=1)=\E(\Delta u_{it}|C=1)$. Instead, for the above DGP to satisfy the assumptions of the method presented in this paper, then two sufficient conditions are i) $\Delta u_{it}\indep D $, ii) $ C_{\Delta u_{t}, u_{t-1}|  d_r=1}= C_{\Delta u_{t}, u_{t-1}| C=1}$. Note that conditions i) and ii) allow for the distribution of the time-varying shock to vary over time -- thus allowing for serial correlation -- as well as for the distribution of $u_{it}$ to be potentially correlated with $\eta_i$, in contrast to \cite{BonhommeSauder}.

The validity of the claims in Example 1 follows the same logic as the proof provided in Appendix A for Example 1 in \cite{CallawayLi2019}. The only difference is that I rely on a copula invariance assumption, while \cite{CallawayLi2019} rely on a copula stability assumption. Furthermore, note that for the method presented in this paper to work, it is sufficient only to impose restrictions on how the distribution of untreated potential outcomes is generated. Nothing is said about how the treated potential outcome is generated.

However, it is highly unlikely that any of Assumptions \hyperref[as:NA]{\ref*{as:NA}}-\hyperref[as:Copula]{\ref*{as:Copula}} hold unconditionally. A large body of literature suggests that, in most empirical applications, these assumptions are more plausibly satisfied after conditioning on pre-treatment observable characteristics \citep[see, e.g.,][]{Heckman1998, Abadie2005}.

In the remainder of this section I, therefore, consider the case where I require Assumptions \hyperref[as:NA]{\ref*{as:NA}}--\hyperref[as:CPTnever]{ \ref*{as:CPTnever}} to hold conditioning on covariates. Specifically, I will distinguish among two scenarios:
\begin{enumerate}
    \item[(i)] Assumptions \hyperref[as:NA]{\ref*{as:NA}} and \hyperref[as:CPTnever]{ \ref*{as:CPTnever}} hold after conditioning on covariates, whereas Assumption \hyperref[as:Copula]{ \ref*{as:Copula}} hold unconditionally;
    \item[(ii)]  Assumptions \hyperref[as:NA]{\ref*{as:NA}}--\hyperref[as:Copula]{ \ref*{as:Copula}} hold after conditioning on covariates.
\end{enumerate}

Let us consider first scenario (i). Then, the only part of \hyperref[Th1]{Theorem \ref*{Th1}} we need to modify to reach identification of the counterfactual distribution is the identification of $F_{\Delta Y(0) |d_r=1 }$, once having accounted for pre-treatment observable characteristics. Rather than obtaining a conditional version of this distribution function, following \cite{CallawayLi2019}, I will generalize the propensity score method proposed by \cite{Firpo2007} to the staggered treatment adoption scenario. 

Despite requiring a parametric specification for $\P_{r,t}(X)$ comes at the additional cost of imposing a distributional form for which there is no guarantee to be correct, this is a commonly used technique that is easy to implement \citep{Abadie2005}: in the first step one obtains an estimator of the propensity score; in a second stage, one plugs the estimated propensity score in the sample counterparts of $F_{Y(0)|d_r=1}$.\footnote{A way to overcome the fact that this method strongly relies on the correct specification of the functional form of $\P_{r,t}(X)$ would be to use a doubly robust estimator for the $F_{Y(0)|d_r=1}$ as the one proposed by \cite{Miller2023}. However, this goes beyond the scope of this paper, whose aim is to propose a method to estimate $F_{Y(0)|d_r=1}$ in a setting characterized by a staggered intervention.} It is worth stressing that the remaining part of  \hyperref[Th1]{Theorem \ref*{Th1}} will continue to be valid thanks to \hyperref[as:Copula]{Assumption \ref*{as:Copula}} holding unconditionally.

\begin{prop}\label{Prop1}
\textit{Under Assumptions 1-4, 6, 7, and unconditional version of Assumptions 5 hold, } $F_{Y_{t}(0)|d_r=1}\left(\tau\right)$ \textit{is identified}: 

\begin{equation*}
         \begin{aligned}
             F_{  Y_{t}\left( 0 \right)|d_r=1}  = & \E \left[ \mathbbm{1} \left[F^{p, -1}_{(Y_t \left(0\right)-Y_{r-\rho-1} \left(0\right)|d_{r}=1)}\left(F_{(Y_t \left(0\right)-Y_{r-\rho-1} \left(0\right)|C=1)}\left(Y_t \left(0\right)-Y_{r-\rho-1} \left(0\right)\right)\right) \right. \right.\\
            & \left. \left. \leq y - F^{-1}_{Y_{r-\rho-1}\left(0 \right) |d_{r}=1} \left(F_{Y_{r-\rho-1}\left(0\right) |C=1} \left(Y_{r-\rho-1}(0)\right)\right)\right]  |C=1 \right]
         \end{aligned} 
         \end{equation*}

         \textit{where}
        \begin{equation}\label{propscoreFirpo}
        F^{p}_{\Delta_{\left[r-\rho-1,t\right]} Y\left(0\right)| d_{r}=1}(\delta)=\E\left[ \frac{C}{p_r} \frac{p_r(x)}{1-p_r(x)} \mathbbm{1} \{ Y_t-Y_{r-\rho-1}  \leq \delta\}\right]        
        \end{equation}
       \textit{which is identified.}
\end{prop}

Again, the result just obtained is almost identical to the one derived in \hyperref[Th1]{Theorem \ref*{Th1}}. The only part that changes is that now  $F_{\Delta Y(0) |d_r=1 }$ is identified by the reweighted distribution in \hyperref[propscoreFirpo]{Eq. (\ref*{propscoreFirpo})} of the change in the untreated potential outcome that occurs for never-treated units. This reweighting procedure is almost identical to the one proposed by \cite{Abadie2005} and \cite{Firpo2007} with some minor changes. For a sketch of the proof of why the result obtained in \hyperref[Th1]{Theorem \ref*{Th1}} is still valid, see Appendix A.

The following example shows the importance of \hyperref[as:NA]{Assumption \ref*{as:NA}} and \hyperref[as:CPTnever]{Assumption \ref*{as:CPTnever}} holding conditionally in the empirical application.

\texttt{Example 2.} Consider the following two-way fixed effects regression for the untreated potential outcome:
\begin{equation*}
       Y_{it}(0)= \alpha_t + \eta_i + X'_{it}\beta+u_{it} 
   \end{equation*}

Where $X$ is a full set of covariates, which can be distributed differently between treated and never-treated units; $u_{it}=\rho u_{i,t-1} + \varepsilon_{it}$, with $\varepsilon \sim WNN$ process and $|\rho|<1$.  For this DGP to satisfy the assumptions of the model presented in this paper, then two sufficient conditions are i) $\Delta u_{it}\indep D | X $, ii) $ C_{\Delta u_{t}, u_{t-1}| d_r=1}= C_{\Delta u_{t}, u_{t-1}| C=1}$. 

The last scenario I consider in this subsection is the case where also \hyperref[as:Copula]{Assumption \ref*{as:Copula}} holds after conditioning on covariates (scenario (ii)). Specifically, the following proposition shows that it is still possible to identify the $F_{Y_{t}(0)|d_r=1}\left(\tau\right)$:

\begin{prop}\label{Prop2}
\textit{Suppose that the random variables $Y_{t-1}(0)$ and $\Delta Y_t(0)$ are continuosly distributed conditionally on $x$, and that this is true for all $x\in\chi$ and both group $r$ or the never-treated units. Then, under Assumptions 1-7,}  $F_{Y_{t}(0)|d_r=1}\left(\tau\right)$ \textit{is identified}: 
\begin{equation*}
         \begin{aligned}
        \P\left(Y_t \left(0\right) \leq y|X=x, d_r=1\right) = & \E\left[ \mathbbm{1} \left[F^{-1}_{\Delta_{\left[r-\rho-1,t\right]} Y\left(0\right)|   X, C=1}\left(F_{\Delta_{r-\rho-1,t} Y\left(0\right)| X, C=1}\left(\Delta_{\left[r-\rho-1, t\right]} Y \left(0\right)\right)| X\right)  \leq y - \right. \right.\\
            & \left. \left. F^{-1}_{Y_{r-\rho-1}\left(0 \right) |X,d_{r}=1} \left(F_{Y_{r-\rho-1}\left(0\right) | X, C=1} \left(Y_{r-\rho-1}(0)\right)|X, \right)\right]  |X=x, C=1 \right]
        \end{aligned} 
        \end{equation*}
        
       \textit{and}
       
    \begin{equation*}
             \P\left(Y_t \left(0\right) \leq y| d_r=1\right) = \int_{\chi}  \P\left(Y_t \left(0\right) \leq y|X=x, d_r=1\right) d F_{X|d_r=1}(x)
        \end{equation*}
\end{prop}

The result obtained in \hyperref[Prop2]{Proposition \ref*{Prop2}} is very close to that obtained above in \hyperref[Th1]{Theorem \ref*{Th1}}. The only difference compared to \hyperref[Th1]{Theorem \ref*{Th1}} is that one needs to compute first conditional distributions of untreated potential outcomes. Next, the researcher has to integrate out covariates from $\P\left(Y_t \left(0\right) \leq y|X=x, d_r=1\right)$ to obtain $ \P\left(Y_t \left(0\right) \leq y| d_r=1\right)$, the unconditional distribution of untreated potential outcome.

\subsubsection{The cohort-time quantile treatment effect}
We can construct various causal parameters once we have identified $F_{Y(0)\mid d_r=1}$. Suppose rank invariance holds. Then, under the same assumptions required for the validity of the results in \hyperref[Th1]{Theorem \ref*{Th1}} or \hyperref[Prop1]{Proposition \ref*{Prop1}}, it is possible to obtain the inverse of $F_{Y(0)\mid d_r=1}$--that is, $F^{-1}_{Y(0)\mid d_r=1}(\tau)$ for each $\tau \in (0,1)$. At this point, the \textit{cohort-time quantile treatment effects} are also identified:

 \begin{equation*}
          QTT_{r,t,\rho}(\tau) = F^{-1}_{Y_{t}(r)|d_r=1}(\tau)- F^{-1}_{  Y_{t}(\infty)|d_r=1}(\tau) \quad \forall  r,t\in\{q,\dots,T\}, t\geq r-\rho.
        \end{equation*}

\sloppy If, instead, one relies on the result obtained in \hyperref[Prop2]{Proposition \ref*{Prop2}}, then after identifying $\mathbb{P}\left(Y_t(0) \leq y \mid X = x, d_r = 1\right)$, one can invert it. Thus, the conditional cohort-time quantile treatment effects are identified.

    \begin{equation*}
         QTT_{r,t,\rho}(\tau; x) = F^{-1}_{Y_{t}(r)|X, d_r=1}(\tau|x)- F^{-1}_{Y_{t}(0)|X, d_r=1}(\tau|x) \quad \forall  r,t\in\{q,\dots,T\} \text{ and } t\geq r-\rho
   \end{equation*}

Alternatively, one can integrate out covariates from $\mathbb{P}\left(Y_t(0) \leq y \mid X = x, d_r = 1\right)$ to obtain $\mathbb{P}\left(Y_t(0) \leq y \mid d_r = 1\right)$, invert it, and then derive:

        \begin{equation*}
          QTT_{r,t,\rho}(\tau) = F^{-1}_{Y_{t}(r)|d_r=1}(\tau)- F^{-1}_{Y_{t}(0)|d_r=1}(\tau) \quad \forall  r,t\in\{q,\dots,T\} \text{ and } t\geq r-\rho
        \end{equation*}

Which is identified.

Once the cohort-time quantile treatment effects are identified, one can aggregate these parameters, for instance, to highlight heterogeneity along specific dimensions (such as how the treatment effects vary with the length of exposure to the treatment). Or, analogously, one can aggregate these causal parameters to summarize the overall treatment effect in one unique parameter. Specifically, I advocate for the researcher to use aggregations schemes of the cohort-time quantile treatment effects similar to those suggested by \cite{Callaway2021}.

Aggregation schemes are particularly crucial when there are many parameters to estimate, and the researcher wants to have a quick tool to summarize results to highlight heterogeneity over time or by treatment cohorts. Note, however, that while in the scenario considered in \cite{Callaway2021}, it made sense to propose a simple weighted average of their treatment effect parameters to summarize the effect of the policy; in principle, in the scenario considered in this paper, one might think of more complicated aggregation schemes to highlight heterogeneity also along the distribution of $Y$. For instance, the researcher might be interested in the policy's overall cumulative effect on the lower tail of the distribution (e.g., evaluate the impact of the introduction of minimum wages on those units that reside in the first two deciles of the income distribution). Nonetheless, to keep things simple and to avoid including an additional dimension of heterogeneity, I will limit myself to generalizing the discussion in \cite{Callaway2021} to the QTT case.

Specifically, in the previous section, we focused on stating under which conditions one can identify the QTT parameters. It may be relevant for a policymaker to understand, for instance, the long-run effects of the policy or to understand whether there exist different patterns according to when a unit is first treated. Following \cite{Callaway2021}, I will propose aggregation schemes of the following form:
\begin{equation}
    \theta(\tau) =\sum_{t=q}^T\sum_{r=q}^Tw(r,t)\cdot QTT_{r,t, \rho}(\tau) \quad r\in\{q,\dots,T\}
\end{equation}

where $w(r,t)$ are the researcher's weighting functions. One can address multiple policy-relevant questions by changing $w(r,t)$, such as: \textit{`How does the QTT vary across groups?'}; \textit{`How does the QTT vary with the length of exposure to the treatment?'}; \textit{`What is the cumulative distributional treatment effect of the policy across all groups until time t?'}; or \textit{`What is the overall impact of the policy'}. For instance, if the researcher is interested in how the QTT varies with the length of exposure to the treatment, let $e = t - r$ denote the event time (i.e., the time elapsed since the unit first received the treatment). One possible aggregation scheme to highlight the heterogeneity of the QTT with respect to event time is:

 \begin{equation*}
        \theta_{exp}^{e}(\tau) =\sum_{r=q}^T\mathbbm{1}\{r+e\leq T\} \P\left(d_r=1|r+e\leq T\right)QTT_{r,r+e, \rho}(\tau) \quad r\in\{q,\dots,T\}
    \end{equation*}

$\theta_{exp}^{e}(\tau)$ represents the effect--on units in the $\tau^{\text{th}}$ quantile of the outcome distribution--of being exposed to the treatment for $e$ periods following its implementation. This parameter is computed using all units that are exactly $e$ periods past the initial treatment date. $\theta_{exp}^{e}(\tau)$ is a parameter commonly studied in event studies in which the researcher aims to understand the dynamic impact of a policy \citep[e.g., see][]{Sun2021, DeChaisemartin2022}.

If, on the other side, the researcher is interested in evaluating the overall impact of the policy on those units that belong to the $\tau^{th}$ quantile, then a straightforward way to obtain an overall effect parameter is the following:
\begin{equation*}
        \theta_{weight}^{o}(\tau) =\frac{1}{\kappa}\sum_{t=q}^T\sum_{r=q}^T\mathbbm{1}\{t\geq r\} \P\left(d_r=1|r\leq T\right)QTT_{r,t, \rho}(\tau) \quad r\in\{q,\dots,T\}
    \end{equation*}

where $\kappa=\sum_{t=q}^T\sum_{r=q}^T\mathbbm{1}\{t\geq r\} \P\left(d_r=1|r\leq T\right)$. $ \theta_{weight}^{o}(\tau)$ is a weighted average of QTTs putting more weight on those QTTs whose size is larger. However, one drawback of $\theta_{weight}^{o}(\tau)$ is that it systematically weights those groups that participate for a longer period in the treatment \citep{Callaway2021}.

\subsubsection{Tests of stochastic dominance rankings}
If rank invariance is unlikely to hold in the context under analysis, one can, instead, perform statistical tests of stochastic dominance or rely on inequality measures. The same reasoning applies when the two distributions cross--i.e., when $QTT_{r,t,\rho}(\tau)$ changes sign as $\tau$ moves from 0 to 1--since policy implications based on these results are sensitive to the choice of quantiles considered. 

For the remainder of this paper, I will focus exclusively on tests of stochastic dominance. Once the counterfactual distribution is identified, one can construct such tests for the treated group $r$ by comparing the distributions of the treated and untreated potential outcomes for the treated--namely, $F_{Y_t(r) \mid d_r = 1}$ and $F_{Y_t(0) \mid d_r = 1}$. Specifically, once the entire counterfactual distribution is known, the researcher can invert it to recover all possible values of $Y_t(0) \mid d_r = 1$, which can then be used to perform stochastic dominance tests.

For example, when considering first-order stochastic dominance, we say that the treated potential outcome $Y_t(r) \mid d_r = 1$ first-order stochastically dominates the untreated potential outcome for the treated $Y_t(0) \mid d_r = 1$ if and only if:
\begin{enumerate}
    \item $\E u(Y_{t}(r)|d_r=1 )\geq \E u(Y_{t}(0)|d_r=1)$ for all $u \in U_1$ with the inequality being strict for some $u$ or
    \item $F_{Y_{t}(r)|d_r=1}(y) \leq F_{Y_{t}(0)|d_r=1}(y)$ for all $Y$ with strict inequality for some values of $Y$, or 
    \item $Y_{t}(r)|d_r=1\geq Y_{t}(0)|d_r=1$ for all points in the support of $Y$.
\end{enumerate}

Again, this is possible as we have identified the entire counterfactual distribution.

%% file: Estimation.tex
\section{Estimation and Inference}\label{Estimation_2}

In the previous sections, we focused on identifying the counterfactual distribution of the untreated potential outcome for the treated group. Once this distribution is identified, we showed that various causal parameters can be constructed. Moreover, when considering the QTT, the aggregation schemes proposed by \cite{Callaway2021} can be naturally extended.

In this section, I propose an approach for estimating these parameters and discuss different methods for conducting valid inference, depending on the estimand of interest.

Suppose the goal is to estimate the parameters defined in \hyperref[qtt]{(\ref*{qtt})}. Using the analogy principle \citep{manski1994analog}, a straightforward way to construct an estimator is to use empirical distribution functions (EDFs), as in \cite{CallawayLi2019}.\footnote{Under the assumptions required in \hyperref[Th1]{Theorem \ref*{Th1}}, and assuming additionally that the support of $Y(a)$ is compact (for $a \in {r, 0}$), \cite{CallawayLi2019} show that the EDF-based estimator is consistent. Moreover, applying a functional central limit theorem, they demonstrate that the estimator converges uniformly to a Gaussian process. These results can be readily extended to the current context, as the only difference lies in the construction of the benchmark, not in the estimator itself.}

If covariates play no role (the case analyzed in \hyperref[Th1]{Theorem \ref*{Th1}}), we can estimate the cohort-time quantile treatment effects by:

            \begin{equation*}
                 \widehat{QTT}_{r,t,\rho}(\tau) = \hat{F}^{-1}_{Y_{t}(r)\mid d_r=1}(\tau)- \hat{F}^{-1}_{Y_{t}(0)\mid d_r=1}(\tau)
            \end{equation*}    
        
where $\widehat{F}^{-1}_{Y_t(r)\mid d_r=1}(\tau)$ is directly estimated from the data. Specifically, the researcher first estimates the empirical distribution function $\widehat{F}_{Y_t(r)\mid d_r=1}(y)$ and then obtains the corresponding quantile for the desired value of $\tau$ by inverting this function. That is:
          \begin{equation*}
                 \hat{F}^{-1}_{Y_{t}(r)\mid d_r=1}(\tau)=inf\{y:\hat{F}_{Y_{t}(r)\mid d_r=1}(y) \geq \tau\}      
            \end{equation*} 

Whereas we can estimate the counterfactual quantile exploiting the result obtained in \hyperref[Th1]{Theorem \ref*{Th1}}, where EDFs and quantile functions will be estimated by their sample analogs:
        \begin{equation*}
            \hat{F}^{-1}_{Y_{t}(0)\mid d_{r}=1}(\tau)=inf\{y:\hat{F}_{Y_{t}(0)\mid d_r=1}(y) \geq \tau\}      
        \end{equation*}
        
        where
        \begin{equation*}
        \begin{aligned}
            \hat{F}_{Y_{t}(0)\mid d_r=1}(y) = & \frac{1}{n_{0}} \sum_{i\in 0} \mathbbm{1} \left[\hat{F}^{-1}_{\Delta_{\left[r-\rho-1,t\right]}Y\mid C=1}\left(\hat{F}_{\Delta_{\left[r-\rho-1,t\right]}Y\mid C=1}\left(\Delta_{\left[r-\rho-1,t\right]}Y\right)\right) \right.\\
            & \left. \leq y - \hat{F}^{-1}_{Y_{r-\rho-1} \mid d_{r}=1} \left(\hat{F}_{Y_{r-\rho-1} \mid C=1} \left(Y_{r-\rho-1}\right)\right)\right] 
        \end{aligned}
        \end{equation*}
where $n_{0}$ represents the number of never-treated units in periods $r-\rho-1$ and $t$.

Regarding statistical inference, the choice of procedure should be guided by the specific features of the context under analysis. For example, when serial correlation is absent or clustered dependence is unlikely, the empirical bootstrap procedure proposed by \cite{CallawayLi2019} is appropriate. They show that this approach yields uniform confidence bands that cover $QTT(\tau)$ with fixed probability for all $\tau \in [\varepsilon, 1 - \varepsilon] \subset (0, 1)$, for some small, positive $\varepsilon$.

Specifically, let $\widehat{QTT}^*(\tau)$ denote a bootstrap estimate of the $QTT$, constructed using the same procedure as above but based on a bootstrap sample. Let $B$ be the total number of bootstrap iterations, indexed by $b = 1, \dots, B$. For each iteration, the researcher computes:
\begin{equation*}
    I^b = sup_{\tau\in \Tau} \hat{\Sigma}(\tau)^{-1/2}\mid \sqrt{n}\left(\widehat{QTT}^b(\tau)-\widehat{QTT}(\tau)\right)\mid
\end{equation*}

Where $\hat{\Sigma}(\tau)^{1/2}=\left(q_{0.75}(\tau)-q_{0.25}(\tau)\right)/\left(z_{0.75}(\tau)-z_{0.25}(\tau)\right)$ represents the interquartile range obtained via bootstrap, divided by the interquartile range of a standard normal random variable. Then one can obtain the $(1-\alpha)$ confidence interval as follows:
\begin{equation*}
    \hat{C}_{QTT(\tau)}= \widehat{QTT}^*(\tau) \pm c_{1-\alpha}^B\hat{\Sigma}(\tau)^{1/2}/\sqrt{n}
\end{equation*}

Where $ c_{1-\alpha}^B$ is the $(1-\alpha)$ quantile of $\{I^b\}_{b=1}^B$.\footnote{For extreme quantile, one can use alternative inference procedures \citep[see,][]{Chernozhukov2016}.} If clustered dependence is likely in the context under analysis, the researcher can adapt the wild cluster bootstrap to the QTT setting with minor modifications. When the number of treated clusters is small, the subcluster wild bootstrap \citep{Mackinnon2018} may be more appropriate. For a comprehensive review of these methods, see \cite{Cameron2015} and \cite{Mackinnon2023}.

In the case considered in \hyperref[Prop1]{Proposition \ref*{Prop1}}, as shown in Appendix A, the only additional term that needs to be estimated is $F^{p,-1}_{\Delta{[r-\rho-1,t]} Y(0) \mid d_r = 1}(\delta)$. This term can be computed by estimating a generalization of the approach proposed by \cite{Firpo2007} and \cite{CallawayLi2019}, as follows:
        \begin{equation*}
            F^{p,-1}_{\Delta{[r-\rho-1,t]} Y(0) \mid d_r = 1}(\delta)=\frac{1}{n_{r,t}} \sum_{i\in r} \frac{C}{p_r} \frac{\hat{p}_r(x_i)}{1-\hat{p}_r(x_i)} \mathbbm{1} \{\Delta Y_{t,r-\rho-1} \leq \delta\}\Bigg/ \frac{1}{n_{r,t}} \sum_{i\in r} \frac{C}{p_r} \frac{\hat{p}_r(x_i)}{1-\hat{p}_r(x_i)} 
        \end{equation*}

Where $n_{r,t}$ is the number of units used to compute $\widehat{QTT}_{r,t,\rho}(\tau)$, $\hat{p}(X)$ represents an estimator of the propensity score, and the last term in the denominator normalizes the weights to sum to 1 in finite samples. This ensures that $\F^{p}(\cdot )$ is indeed a distribution function. It is worth stressing that, as both \cite{Firpo2007} and \cite{CallawayLi2019} show, one can estimate the propensity score parametrically and non-parametrically.

As for the case considered in \hyperref[Prop2]{Proposition \ref*{Prop2}}, it can be computationally demanding. Even if estimated non-parametrically, it requires estimation of five conditional distributions, which may be infeasible in many empirical applications--particularly when the dimension of $\pmb{X}$ is large and the sample size $n$ is small.

To address this issue, one can either estimate the relevant quantities using quantile regression or, when covariates are discrete, apply the method proposed by \cite{Callawayetal2018}. For example, \cite{Santangelo} employs quantile regression to estimate the Change-in-Changes model developed by \cite{AtheyImb2006} in a setting with covariates.

Alternatively, following \cite{Callawayetal2018}, one can estimate $\hat{F}_{Y_t(r) \mid X, d_r = 1}$ using the empirical distribution function for each possible value of $x$, and then invert it. Similarly, $\hat{F}_{Y_t(0) \mid X, d_r = 1}$ can be estimated for each value of $x$ by applying the identification result in \hyperref[Prop2]{Proposition \ref*{Prop2}}. That is:

\begin{equation*}
        \begin{aligned}
            \hat{F}_{Y_{t}(0)\mid X=x, d_r=1}(y) = & \frac{1}{n_{0,x}} \sum_{i\in 0} \mathbbm{1} \left[\left(\Delta_{\left[r-\rho-1,t\right]}Y\right) \right.\\
            & \left. \leq y - \hat{F}^{-1}_{Y_{r-\rho-1} \mid X=x, d_{r}=1} \left(\hat{F}_{Y_{r-\rho-1} \mid X=x, C=1} \left(Y_{r-\rho-1}\right)\right)\mid  X=x\right] 
        \end{aligned}
        \end{equation*}
Again, one can use empirical distribution functions to estimate each of the quantities in the formula above. Once obtained an estimate of $\hat{F}_{Y_{t}(r)\mid X, d_r=1}(\tau\mid x)$ and $\hat{F}_{Y_{t}(0)\mid X, d_r=1}(\tau\mid x)$, one can obtain an estimator of the conditional QTT as: 
\begin{equation*}
         QTT_{r,t,\rho}(\tau; x) = F^{-1}_{Y_{t}(r)\mid X, d_r=1}(\tau\mid x)- F^{-1}_{Y_{t}(0)\mid X, d_r=1}(\tau\mid x) \quad \forall  r,t\in\{q,\dots,T\} \text{ and } t\geq r-\rho
   \end{equation*}

for each $\tau \in \left(0,1\right)$ and for each $x \in \chi$.

Once an estimator for the QTTs is obtained, the researcher can aggregate these parameters using the aggregation schemes outlined in the previous section. The corresponding weighting functions can be estimated using their sample analogs, following the analogy principle \citep{manski1994analog}.

If, on the other hand, the researcher is interested in performing tests of stochastic dominance (SD) rankings, one can perform stochastic dominance tests based on the generalized Kolmogorov-Smirnov statistics proposed by \cite{Maasoumi2005}. Once estimators for $\hat{F}_{Y_t(r) \mid d_r = 1}$ and $\hat{F}_{Y_t(0) \mid d_r = 1}$ are obtained--leveraging the identification results from \hyperref[Th1]{Theorem \ref*{Th1}}, or from \hyperref[Prop1]{Proposition \ref*{Prop1}} or \hyperref[Prop2]{Proposition \ref*{Prop2}}, depending on the setting under analysis--tests for first-order or second-order stochastic dominance (FSD or SSD) can be constructed as follows:

\begin{equation*}
    d_{r,t, \rho} = \sqrt{\frac{n_{r,t} \cdot n_{0,t}}{n_{r,t}+n_{0,t}}} \text{min sup}\left(\hat{F}_{Y_{t}(r)\mid  d_r=1}(y) -\hat{F}_{Y_{t}(0)\mid d_r=1}(y)\right)
\end{equation*}
\begin{equation*}
    s_{r,t, \rho} = \sqrt{\frac{n_{r,t} \cdot n_{0,t}}{n_{r,t}+n_{0,t}}} \text{min sup}\int_{-\infty}^y\left(\hat{F}_{Y_{t}(r)\mid  d_r=1}(z) - \hat{F}_{Y_{t}(0)\mid d_r=1(z)}\right) dz
\end{equation*}

where $n_{r,t}$ and $n_{0,t}$ represent the sample sizes used to estimate $\hat{F}_{Y_{t}(r)\mid  d_r=1}$ and $\hat{F}_{Y_{t}(0)\mid d_r=1}$, respectively. 

Following \cite{Maasoumi2019}, one can construct a test for $iid$ samples based on the pair bootstrap. This procedure allows the researcher to obtain the probability of any of the two SD tests to fall in a specific interval, as well as the $p-$value. For instance, if $\P(d\leq 0)$ is large (e.g., above $.90$) and $d$ is non-positive, then we can claim that $Y_{t}(r)\mid  d_r=1$ FSD $Y_{t}(0)\mid d_r=1$ with a high degree of statistical confidence \citep{Maasoumi2019}.

More sophisticated tests can be developed depending on the specific context. For a comprehensive survey of possible testing procedures within this framework, see \cite{Maasoumi2001}. As emphasized earlier, the choice of test should be guided by the characteristics of the empirical setting to appropriately assess whether the distribution of the treated potential outcome at time $t$ for the group first treated in period $r$ stochastically dominates the corresponding counterfactual distribution. A detailed discussion of these testing strategies, however, is beyond the scope of this paper.

%% file: Simulation.tex
\section{Monte-Carlo Simulations}\label{Simulation}

\subsection{Monte-Carlo Simulations -- varying \texorpdfstring{$N$}{N} and \texorpdfstring{$T$}{T}}
In this section, I examine the finite-sample properties of the estimators for the distribution of the untreated potential outcome introduced in the previous section. Since the identification results presented in \hyperref[Identification]{Section \ref*{Identification}} do not depend on the specific causal parameter of interest, I focus, for brevity, on the case where the researcher aims to estimate the cohort-time quantile treatment effect. In particular, because identification does not require specifying the outcome distribution for the treated group, I restrict the attention to assessing the performance of the proposed methods in recovering $F_{Y_t(0) \mid d_r = 1}$.\footnote{It is worth noting that the rank invariance assumption may plausibly hold in the models considered here, as no assumptions are made about $Y_t(r) \mid d_r = 1$.}

In contrast to most of the existing literature, I assess the performance of the proposed method in three scenarios: (i) when the DGP for the untreated potential outcome follows a TWFE regression without covariates \citep[as in][]{Callawayetal2018, CallawayLi2019, Miller2023}; (ii) when the linear TWFE regression includes covariates \citep[as in][]{Firpo2007, Callaway2021}; and (iii) when $Y_{it}(0)$ is generated by a panel quantile regression.

Throughout this section, without any loss of generality, I will assume that there is no treatment anticipation (i.e., $\rho=0$), that the policy is implemented starting from the second period onward, and that panel data are available. Let $T$ denote the maximum number of available time periods, $q=2$ the first time a policy is implemented, $r=\{2,\dots, T\}$ the first time a unit is treated, and $n$ the sample size. I allow both $n$ and $T$ to vary in the following Monte Carlo exercises. I will compare the proposed estimators' performance in terms of average bias and root-mean-squared error (RMSE). For completeness, when considering the DGPs in which covariates play a role, I also consider the unconditional QTT parameter, which can be obtained using the identification result in \hyperref[Th1]{Theorem \ref*{Th1}}. \\

\noindent \textit{DGP 1}: The first DGP considered is the following:

\begin{equation}
\label{DGP1}
       Y_{it}(0)= \alpha_t + \eta_i +u_{it}
   \end{equation}

Where $Y_{it}(0)$ denotes the untreated potential outcome, $\alpha_t$ are time fixed-effects, $u_{it}$ represents the time-varying random shock, and $\eta_i$ represents the time-invariant unobserved heterogeneity, which can be distributed differently across treated cohorts and never-treated units (i.e., I allow for selection on unobservables). \hyperref[DGP1]{Equation~(\ref*{DGP1})} is similar to DGPs considered in \cite{Callawayetal2018, CallawayLi2019, Miller2023}.

Specifically, in \textit{DGP 1}, I set  $\alpha_t =t$, $\eta |d_r=1 \sim N(r,1)$ with $r=\{2,\dots, T\}$, and $u_t\sim N(0,1)$ in all periods. Implicitly assumed is that $T=R$, where $R$ represents the last time period in which a unit can undergo the policy (i.e., there is a positive probability of being treated in every period starting from the second period onward). Since covariates do not play a role in this DGP, I set the probability of being first treated in period $r$, $\P(d_r=1)$, to $1/T$. Then $Y_{it}(0)\sim N(\alpha_t + r, 2)$. Exploiting the fact that for a generic random variable (R.V.) $X\sim N(\mu, \sigma^2)$, the $\tau$th quantile is given by $F_X^{-1}(\tau)=\mu+ \sigma \Phi^{-1}\left(\tau\right)$--where $\Phi(\cdot)$ represents the standard normal CDF. Then the $\tau$th quantile at time $t$ for units belonging to cohort $r$ is equal to $F_{Y_{t}(0)|d_r=1}^{-1}(\tau)= \left(\alpha_t +r\right) + \sqrt{2}\Phi^{-1}\left(\tau\right)$.\\

\noindent \textit{DGP 2}: The second DGP mimics the TWFE regression with covariates used by \cite{Callaway2021} to study the finite-sample properties of the estimators of the group-time average treatment effects. Specifically, in contrast to \textit{DGP 1}, it also allows for the selection on observable characteristics. I will consider the case where there is only one covariate $X \sim N(0,1)$ for simplicity. Since in this DGP, selection on observables is allowed, I set the probability to belong to cohort $r$ to be a function of observable pre-treatment characteristics \citep[as in][]{Callaway2021}:

\begin{equation}
\label{GeneralizedPS}
         \P(d_r=1|X)=\frac{exp(X'\gamma_r)}{1+\sum_r exp(X'\gamma_r)} 
    \end{equation}
    
where $\gamma_r=0.5r/T$. The following model generates the untreated potential outcome:
   \begin{equation}
       \label{DGP2}
       Y_{it}(0)=  \alpha_t + \eta_i + X_{it}+u_{it} 
   \end{equation}

Compared to \textit{DGP 1}, the only difference is that now I also allow selection on observable characteristics. I still assume $\alpha_t =t$, $\eta |d_r=1 \sim N(r,1)$ with $r=\{2,\dots, T\}$, and $u_t\sim N(0,1)$. In this case, since $Y_{it}(0)\sim N(\alpha_t + r, 3)$, the (population) $\tau$th quantile of $Y_{t}(0)|d_r=1$ is equal to $ F_{Y_{t}(0)|d_r=1}^{-1}(\tau)= \left(\alpha_t +r\right) + \sqrt{3}\Phi^{-1}\left(\tau\right)$. 

In Appendix C, I also present the results for \textit{DGP 2}, but instead of assuming $\alpha_t=t$, I examine the estimator's performance in the presence of a quadratic trend: $\alpha_t=t+t^2$.\\

\noindent \textit{DGP 3}: The third DGP generates $Y_{it}(0)$ by a panel quantile regression. Specifically, the following DGP is a re-adaptation of the DGP presented in \cite{Machado}:
   
   \begin{equation}
   \label{DGP3}
       Y_{it}(0)=  \alpha_t + \eta_i + X_{it}+\left(1+X_{it}\right)u_{it}
   \end{equation}
where the probability of being treated is still given by \hyperref[GeneralizedPS]{(\ref*{GeneralizedPS})}, $u_t\sim N(0,1)$, and $\alpha_t =t$. Whereas now $\eta|d_r=1 \sim Z_r$, $\gamma_r=\frac{r}{4T}$, $X_{it} = \frac{1}{8} \chi_{it}$, with $Z_r= r + T$, $T \sim \chi^2_{(1)}$ and $\chi_{it} \sim \chi^2_{(1)}$. 

While in \textit{DGP 1} and \textit{DGP 2}, it was possible to obtain an analytical formula to compute the population quantiles of $Y_{t}(0)|d_r=1$, this is not the case for \textit{DGP 3}. The distribution of $Y_{it}(0)$ is difficult to retrieve. For this reason, I approximate the population quantiles with those obtained by simulating \hyperref[DGP3]{(\ref*{DGP3})} using $1$ million observations. \\

All DGPs presented in this section satisfy the assumptions needed for identification.\footnote{Sufficient conditions for the distributional PT assumption and the copula invariance assumption to hold in linear models is that the error term is $iid$ in each period. In the nonlinear DGP, this is complicated by the term $(1+X_{it})u_{it}$. However, even in this case, one can show that the distributional PT holds conditionally and the copula invariance holds unconditionally as long as $X$ and $u$ are independent of treatment assignment.} Specifically, in \textit{DGP 1}, the distributional parallel trends and the copula invariance assumptions hold unconditionally. Whereas, in \textit{DGP 2} and \textit{DGP 3}, the distributional PT assumption holds conditionally on $X$, whereas the copula assumption still holds unconditionally.\footnote{The reason why the copula invariance still holds unconditionally is that none of the random variables in the right-hand side of the equation are distributed conditionally on $X$. As pointed out in the previous section, estimation may become demanding when the copula invariance holds conditionally.}

In this subsection, I consider the following scenarios: i) $T=R=4$ and ii) $T=R=10$. I allow the sample size to vary in each setup, $n=\{100, 1000\}$. All results are based on $2,000$ Monte Carlo simulations, and I report the results for the $.25$, $.50$, and $.75$ quantiles. To save space only results for $F_{Y_{2}(0)|d_2=1}^{-1}(\tau)$ are shown in the main text. In Appendix C, I present the results obtained for scenario i) for all parameter estimators when the .5 quantile is considered. Based on the theoretical results obtained in \cite{CallawayLi2019} and \cite{Callaway2021}, the estimators are expected to perform well when $n$ is large relative to $R$ and $T$. What happens when $T$ (and consequently $R$) grows is unclear, as the number of units in a given cohort can be very small.

\begin{table}[htbp]
\caption{\footnotesize Monte Carlo Results for $F_{Y_{2}(0)|d_2=1, \rho=0}^{-1}(\tau)$. Setup with $T=R=4$.}
\label{montecarlo1}
\resizebox{\textwidth}{!}{
\begin{tabular}{l*{11}{c}}
\toprule
\small
\vspace*{0.0002cm}	\\																								
\vspace*{0.0002cm}	\\																								
				&	&	 & \multicolumn{3}{c}{\textbf{0.25}} 					&\multicolumn{3}{c}{\textbf{0.50}} 				& \multicolumn{3}{c}{\textbf{0.75}} 			\\							
	\cmidrule(rl){4-6} \cmidrule(rl){7-9} \cmidrule(rl){10-12} 																								
\vspace*{0.0025cm}	\\																								
		&	&	\textbf{n}	&	$F_{Y_{t}(0)|d_r=1}^{-1}$ 	&	Bias	&	Root MSE 	&		$F_{Y_{t}(0)|d_r=1}^{-1}$ 	&	Bias	&	Root MSE 	&		$F_{Y_{t}(0)|d_r=1}^{-1}$ 	&	Bias	&	Root MSE 	\\
  \vspace*{0.0000001cm}	\\																								
\cline{2-12}																									
  \vspace*{0.0025cm}	\\																								
\multirow{1}{4em}{\textbf{DGP 1}}	&	\textbf{Unc}	&	100	&	3.046	&	0.112	&	0.525	&		4.000	&	0.097	&	0.489	&		4.954	&	0.122	&	0.509	\\
\multirow{2}{4em}{\textbf{DGP 2}}	&	\textbf{Unc}	&	100	&	2.832	&	0.331	&	0.695	&		4.000	&	0.341	&	0.684	&		5.168	&	0.391	&	0.714	\\
	&	\textbf{Cond}	&	100	&	2.832	&	0.086	&	0.531	&		4.000	&	0.091	&	0.515	&		5.168	&	0.146	&	0.541	\\
\multirow{2}{4em}{\textbf{DGP 3}}	&	\textbf{Unc}	&	100	& 3.929 & 0.178 & 0.550	&		4.839 & 0.244 & 0.583	&		5.925 & 0.361 & 0.721	\\
	&	\textbf{Cond}	&	100	&	3.929 & 0.114 & 0.571	&		4.839 & 0.203 & 0.590	&		5.925 & 0.335 & 0.730	\\    
\vspace*{0.005cm}																									\\
\multirow{1}{4em}{\textbf{DGP 1}}	&	\textbf{Unc}	&	1000	&	3.046	&	0.011	&	0.152	&		4.000	&	0.007	&	0.15	&		4.954	&	0.013	&	0.157	\\
\multirow{2}{4em}{\textbf{DGP 2}}	&	\textbf{Unc}	&	1000	&	2.832	&	0.258	&	0.314	&		4.000	&	0.25	&	0.304	&		5.168	&	0.257	&	0.315	\\
	&	\textbf{Cond}	&	1000	&	2.832	&	0.018	&	0.154	&		4.000	&	0.011	&	0.147	&		5.168	&	0.018	&	0.158	\\
\multirow{2}{4em}{\textbf{DGP 3}}	&	\textbf{Unc}	&	1000	&	3.929 & 0.061 & 0.167	&		4.839 & 0.085 & 0.180	&		5.925 & 0.132 & 0.230	\\
	&	\textbf{Cond}	&	1000	&	3.929 &-0.005 & 0.163	&	4.839 & 0.040 & 0.169	&		5.925 & 0.101 & 0.218	\\    
\bottomrule
\end{tabular}}
\floatfoot{\footnotesize{\textbf{Notes}: This table reports the Monte Carlo results for the estimator of the parameter $F_{Y_{2}(0)|d_2=1, \rho=0}^{-1}(\tau)$ in the setup with $T=R=4$. Results are reported for the quantiles $.25$, $.50$, $.75$. Each Monte Carlo simulation uses $2,000$ bootstrap replications. Rows labeled `DGP 1' report the results obtained for the DGP in \hyperref[DGP1]{Eq. (\ref*{DGP1})}, rows labeled `DGP 2' present the results obtained for the DGP in \hyperref[DGP2]{Eq. (\ref*{DGP2})}, and rows labeled `DGP 3' for the DGP in \hyperref[DGP3]{Eq. (\ref*{DGP3})}. Rows labeled `UNC' use the estimator based on the unconditional distributional PT assumption that ignores covariates. Finally, $F_{Y_{t}(0)|d_r=1}^{-1}$ represents the true population parameter, whereas `Bias' and `RMSE' stand for average (simulated) bias, and root-mean-squared, respectively.}}
\end{table}  

\hyperref[montecarlo1]{Table~\ref*{montecarlo1}} reports the simulation results for the estimators of $F_{Y_{2}(0)|d_2=1}^{-1}(\tau)$ in the scenario where $T=R=4$, across all three DGPs presented. As expected, when the number of units per cohort is small ($n=100$), the performance of the method presented is poor--this is especially true for the nonlinear DGP. As the number of units per cohort increases, the performance improves substantially in terms of bias and RMSE. Interestingly, the largest gain--in terms of bias reduction--is observed when estimating the parameters of \textit{DGP 3}. This is likely because, when $n$ is small, the proposed method `struggles more' to recover the true counterfactual distribution under a nonlinear DGP compared to a linear one, as fewer observations are available in each quantile. As $n$ grows, however, the number of observations per quantile increases, leading to substantial improvements in the method's performance.

Moreover, as the theory predicts, ignoring the role of covariates substantially reduces the estimator's performance, leading to unreliable inference. 

Similar results are found for the other cohort-time quantile treatment effect estimators. To save space, these results are omitted from the main text. Tables \hyperref[MC_dgp1_appendix]{\ref*{MC_dgp1_appendix}}, \hyperref[MC_dgp2_appendix]{\ref*{MC_dgp2_appendix}}, and \hyperref[MC_dgp3_appendix]{\ref*{MC_dgp3_appendix}} in Appendix C report the full set of results for the median--obtained under the scenario with $T=4$--for \textit{DGP 1}, \textit{DGP 2}, and \textit{DGP 3}, respectively. All additional results are available upon request.

In \hyperref[montecarlo_nonlinear_trends]{Table~\ref*{montecarlo_nonlinear_trends}} in Appendix C, I present the results for the estimator of $F_{Y_{2}(0)|d_2=1}^{-1}(\tau)$ under the scenario with $T=R=4$ for \textit{DGP 2}, which now includes a quadratic trend instead of a linear one.

The results indicate that the estimator's performance--both in terms of bias and RMSE--is nearly identical to the case where $\alpha_t=t$ shown in \hyperref[montecarlo1]{Table \ref*{montecarlo1}}. This result is expected since, for \hyperref[as:CPTnever]{Assumption~\ref{as:CPTnever}} to hold, the key requirement is that trends in untreated potential outcomes be independent of treatment assignment. 

It is worth stressing that we would reach a similar conclusion to the one described above (i.e., the presence of non-linear trends) if, instead of requiring the error term to be independent and identically distributed, we allowed the errors to be heteroskedastic (or within-cluster correlation to be present). As long as the structure of the heteroskedasticity/clustering is independent of treatment assignment, \hyperref[as:CPTnever]{Assumption~\ref{as:CPTnever}} and/or \hyperref[as:Copula]{Assumption~\ref*{as:Copula}} still hold. Indeed, a sufficient condition for these two assumptions to be satisfied is that, even if clustering is present, the heteroskedasticity/within-cluster correlation remains unaffected by treatment assignment. Suppose, absurdly, that we know the skedastic function (e.g., $exp\{X\beta\}$ with $\beta \in \mathbb{R}$) and that $Y_{it}(0)$ is generated according \hyperref[DGP2]{(\ref*{DGP2})}. A sufficient condition for the Distributional PT assumption to hold is that the skedastic function has a functional form that remains unchanged between treated groups and never-treated units. The Copula Invariance Assumption, on the other hand, requires that the dependence structure between the pre-treatment level of the error term, $u_{ir}$, and its change after the introduction of the policy, $\Delta_{r-1,t}u$ must remain unaffected by treatment assignment.

\begin{table}[htbp]
\caption{\footnotesize Monte Carlo Results for $F_{Y_{2}(0)|d_2=1, \rho=0}^{-1}(\tau)$. Setup with $T=R=10$.}
\label{montecarlo2}
\resizebox{\textwidth}{!}{
\begin{tabular}{l*{11}{c}}
\toprule
\small
\vspace*{0.0002cm}	\\																								
				&	&	 & \multicolumn{3}{c}{\textbf{0.25}} 					&\multicolumn{3}{c}{\textbf{0.50}} 				& \multicolumn{3}{c}{\textbf{0.75}} 			\\							
	\cmidrule(rl){4-6} \cmidrule(rl){7-9} \cmidrule(rl){10-12} 																								
\vspace*{0.0025cm}	\\																								
		&	&	\textbf{n}	&	$F_{Y_{t}(0)|d_r=1}^{-1}$ 	&	Bias	&	Root MSE 	&		$F_{Y_{t}(0)|d_r=1}^{-1}$ 	&	Bias	&	Root MSE 	&		$F_{Y_{t}(0)|d_r=1}^{-1}$ 	&	Bias	&	Root MSE 	\\
  \vspace*{0.0000001cm}	\\																								
\cline{2-12}																									
  \vspace*{0.0025cm}	\\																								
\multirow{1}{4em}{\textbf{DGP 1}}	&	\textbf{Unc}	&	100	&	3.046	&	0.047	&	0.858	&		4.000	&	0.073	&	0.864	&		4.954	&	0.137	&	0.895	\\
\multirow{2}{4em}{\textbf{DGP 2}}	&	\textbf{Unc}	&	100	&	2.832	&	0.114	&	1.087	&		4.000	&	0.196	&	1.082	&		5.168	&	0.306	&	1.091	\\
	&	\textbf{Cond}	&	100	&	2.832	&	-0.02	&	0.927	&		4.000	&	0.085	&	0.911	&		5.168	&	0.173	&	0.918	\\
\multirow{2}{4em}{\textbf{DGP 3}}	&	\textbf{Unc}	&	100	&	3.935&0.091&0.925	&		4.861&0.209&0.996	&		5.971&0.377&1.337	\\
	&	\textbf{Cond}	&	100	&	3.935&0.051&0.997	&		4.861&0.189&1.053	&		5.971&0.354&1.371	\\
\vspace*{0.005cm}																									\\
\multirow{1}{4em}{\textbf{DGP 1}}	&	\textbf{Unc}	&	1000	&	3.046	&	0.013	&	0.245	&		4.000	&	0.005	&	0.232	&		4.954	&	0.013	&	0.251	\\
\multirow{2}{4em}{\textbf{DGP 2}}	&	\textbf{Unc}	&	1000	&	2.832	&	0.107	&	0.31	&		4.000	&	0.111	&	0.315	&		5.168	&	0.125	&	0.319	\\
	&	\textbf{Cond}	&	1000	&	2.832	&	0.008	&	0.247	&		4.000	&	0.012	&	0.245	&		5.168	&	0.026	&	0.249	\\	
\multirow{2}{4em}{\textbf{DGP 3}}	&	\textbf{Unc}	&	1000	&	3.935&0.044&0.261	&		4.861&0.069&0.274	&		5.971&0.100&0.327	\\
	&	\textbf{Cond}	&	1000	&	3.935&-0.005&0.261 &		4.861&0.034&0.269	&		5.971&0.078&0.319	\\	
\bottomrule
\end{tabular}}
\floatfoot{\footnotesize{\textbf{Notes}: This table reports the Monte Carlo results for the estimator of the parameter $F_{Y_{2}(0)|d_2=1, \rho=0}^{-1}(\tau)$ in the setup with $T=R=10$. Results are reported for the quantiles $.25$, $.50$, $.75$. Each Monte Carlo simulation uses $2,000$ bootstrap replications. Rows labeled `DGP 1' report the results obtained for the DGP in \hyperref[DGP1]{Eq. (\ref*{DGP1})}, rows labeled `DGP 2' present the results obtained for the DGP in \hyperref[DGP2]{Eq. (\ref*{DGP2})}, and rows labeled `DGP 3' for the DGP in \hyperref[DGP3]{Eq. (\ref*{DGP3})}. Rows labeled `UNC' use the estimator based on the unconditional distributional PT assumption that ignores covariates. Finally, $F_{Y_{t}(0)|d_r=1}^{-1}$ represents the true population parameter, whereas `Bias' and `RMSE' stand for average (simulated) bias, and root-mean-squared, respectively.}}
\end{table}  

Regarding the scenario with $T=R=10$, results are reported in \hyperref[montecarlo2]{Table~\ref*{montecarlo2}} . Again, also in this case, I report the results for the estimators of $F_{Y_{2}(0)|d_2=1}^{-1}(\tau)$ for the three DGPs. As with $T=R=4$, when the number of units per cohort is small, the method's performance is poor, but as $n$ increases, the method's performance improves substantially. 

Interestingly, the average bias for the estimators of the parameters of all DGPs now appears to be smaller in magnitude than those shown in  \hyperref[montecarlo1]{Table~\ref*{montecarlo1}} (except for $\tau=.75$ in the scenario with $n=100$, where the magnitude is larger). However, for almost all cases, the RMSEs are larger than those shown in \hyperref[montecarlo1]{Table~\ref*{montecarlo1}}, so the parameters are now estimated imprecisely. The fact that the RMSEs are larger than before is probably because the number of units per cohort is much smaller. 

Lastly, in the scenario with $T=R=10$ and small $n$, the parameters of \textit{DGP 2} and \textit{DGP 3} could not be estimated in many simulations due to the low number of treated and/or never-treated units (results not shown). 

So, overall, the performance of the estimators of the parameters of \textit{DGP 2} and \textit{DGP 3} does not necessarily improve compared to the scenario in which $T=R=4$.

All the results shown are aligned with theoretical predictions and consistent with the simulation results in \cite{Callawayetal2018, CallawayLi2019, Callaway2021}. The estimator's performance is relatively poor for small $n$ but improves substantially as $n$ increases. Specifically, all Monte Carlo results support the $\sqrt{n}$-consistency of the estimator for $F_{Y_{t}(0)|d_r=1, \rho}^{-1}(\tau)$. Indeed, in both \hyperref[montecarlo1]{Table~\ref*{montecarlo1}} and \hyperref[montecarlo1]{Table~\ref*{montecarlo2}}, the RMSE decreases approximately by a factor of $1/\sqrt{10}$ when the sample size increases from $n=100$ to $n=1000$, across all DGPs and regardless of the conditioning.

Lastly, in line with \cite{Callawayetal2018}, I also found that, for all the DGPs, the method's power is relatively larger for the $.50$ quantile.

\subsection{Monte-Carlo Simulations -- violations of the main identifying assumptions}

In this subsection, I analyze the performance of the proposed estimator in generating $F_{Y_t(0)|d_r=1}^{-1}\left(\tau \right)$ when the untreated potential outcome follows a TWFE regression with covariates, similar to \hyperref[DGP2]{(\ref*{DGP2})}. However, in contrast to \textit{DGP 2}, I now assess the impact of small deviations in:
\begin{itemize}
    \item The Distributional Parallel Trends assumption (while the Copula Invariance assumption continues to hold), and
    \item The Copula Invariance assumption (while maintaining the Distributional Parallel Trends assumption).
\end{itemize}

As in the previous subsection, I assume no treatment anticipation, that the policy is implemented starting from the second period onward, and that panel data are available. I set $n=1000$ and $T=R=4$. Since covariates play a role, for completeness, I also evaluate the performance when considering the unconditional parameter. Further, as before, I assume that $X\sim N(0,1)$, and the probability of belonging to cohort $r$ follows Eq. \hyperref[GeneralizedPS]{(\ref*{GeneralizedPS})}, with $\gamma_r=0.5r/T$.\\

\noindent \textit{DGP 4}: In this DGP, I assess the performance of the proposed estimator in situations where \hyperref[as:Copula]{Assumption~\ref*{as:Copula}} still holds unconditionally, as in \textit{DGP 2}, but now I analyze the effects of small deviations from \hyperref[as:CPTnever]{Assumption~\ref*{as:CPTnever}}. To this end, I assume the untreated potential outcome follows the following TWFE regression:

\begin{equation}
     \label{DGP4}
       Y_{it}(0)=  \alpha_{t,d} + \eta_i +  X_{it}+u_{it} 
\end{equation}

where I still assume $\eta |d_r=1 \sim N(r,1)$ with $r=\{2,\dots, 4\}$, and $u_t\sim N(0,1)$. However, now $\alpha_{t,d}=t\left(1+\bar \varepsilon d\right)$, where $d=0$ for never-treated units and $d=1$ for treated cohorts, and $\bar \varepsilon$ represents the degree of violation of \hyperref[as:CPTnever]{Assumption~\ref*{as:CPTnever}} allowed. When $\bar \varepsilon=0$, the Conditional Distributional Parallel Trends hold. When $\bar \varepsilon \neq 0$, \hyperref[as:CPTnever]{Assumption~\ref*{as:CPTnever}} is violated. The larger $\bar \varepsilon d$ is, the greater the deviation.   

By allowing $\alpha_{t,d}$ to vary, we introduce violations of the Conditional Distributional Parallel Trends between treated and never-treated units. Note that no distinction in violations between treated cohorts is considered, as this would complicate the analysis.

In this case, since $Y_{it}(0)\sim N\left(t\left(1+\bar \varepsilon d\right) + r, 3\right)$, the (population) $\tau$th quantile of $Y_{t}(0)|d_r=1$ is equal to $ F_{Y_{t}(0)|d_r=1}^{-1}(\tau)= t\left(1+\bar \varepsilon d\right) +r + \sqrt{3}\Phi^{-1}\left(\tau\right)$. 

\begin{table}[htbp]
\caption{\footnotesize Monte Carlo Results, DGP 4, for $F_{Y_{2}(0)|d_2=1, \rho=0}^{-1}(\tau)$ -- Violation Cond. Distributional PT.}
\label{montecarlo3}
\resizebox{\textwidth}{!}{
\begin{tabular}{l*{10}{c}}
\toprule
\small
\vspace*{0.0002cm}	\\																								
\vspace*{0.0002cm}	\\																								
				&	& \multicolumn{3}{c}{\textbf{0.25}} 					&\multicolumn{3}{c}{\textbf{0.50}} 				& \multicolumn{3}{c}{\textbf{0.75}} 			\\							
	\cmidrule(rl){3-5} \cmidrule(rl){6-8} \cmidrule(rl){9-11} 																								
\vspace*{0.0025cm}	\\																								
	$\bar \varepsilon$	&	&	$F_{Y_{t}(0)|d_r=1}^{-1}$ 	&	Bias	&	Root MSE 	&		$F_{Y_{t}(0)|d_r=1}^{-1}$ 	&	Bias	&	Root MSE 	&		$F_{Y_{t}(0)|d_r=1}^{-1}$ 	&	Bias	&	Root MSE 	\\
  \vspace*{0.0000001cm}	\\																								
\cline{1-11}																									
  \vspace*{0.0025cm}	\\																								
	
\multirow{2}{4em}{\textbf{0.00}}	&	\textbf{Unc}	& 2.832	&	0.257	&	0.314	&	4.000	&	0.253	&	0.307	&	5.168	&	0.253	&	0.311	\\
                                    &	\textbf{Cond}	& 2.832	&	0.017	&	0.155	&	4.000	&	0.012	&	0.15	&	5.168	&	0.013	&	0.158	\\
																					
\multirow{2}{4em}{\textbf{0.05}}	&	\textbf{Unc}	&	2.932	&	0.207	&	0.274	&	4.100	&	0.203	&	0.27	&	5.268	&	0.206	&	0.274	\\
                                    &	\textbf{Cond}	&	2.932	&	-0.033	&	0.158	&	4.100	&	-0.037	&	0.158	&	5.268	&	-0.033	&	0.162	\\
																					
\multirow{2}{4em}{\textbf{0.10}}   &	\textbf{Unc}	&	3.032	&	0.157	&	0.242	&	4.200	&	0.155	&	0.231	&	5.368	&	0.151	&	0.239	\\
                                   &	\textbf{Cond}	&	3.032	&	-0.083	&	0.177	&	4.200	&	-0.086	&	0.166	&	5.368	&	-0.089	&	0.183	\\
																					
\multirow{2}{4em}{\textbf{0.50}}   &	\textbf{Unc}	&	3.832	&	-0.239	&	0.297	&	5.000	&	-0.248	&	0.304	&	6.168	&	-0.252	&	0.312	\\
                                   &	\textbf{Cond}	&	3.832	&	-0.48	&	0.504	&	5.000	&	-0.488	&	0.51	&	6.168	&	-0.492	&	0.516	\\
\bottomrule
\end{tabular}}
\floatfoot{\footnotesize{\textbf{Notes}: This table reports the Monte Carlo results of the DGP in \hyperref[DGP4]{Eq. (\ref*{DGP4})} for the estimator of the parameter $F_{Y_{2}(0)|d_2=1, \rho=0}^{-1}(\tau)$ in the setup with $T=R=4$ and violations of the Distributional PT assumption. Results are reported for the quantiles $.25$, $.50$, $.75$. Each Monte Carlo simulation uses $2,000$ bootstrap replications. Rows labeled `UNC' use the estimator based on the unconditional distributional PT assumption that ignores covariates. Here, $\bar \varepsilon$ controls whether the Distributional PT is violated. When $\bar \varepsilon=0$, \hyperref[as:CPTnever]{Assumption~\ref*{as:CPTnever}} is not violated. When $\bar \varepsilon \neq 0$, \hyperref[as:CPTnever]{Assumption~\ref*{as:CPTnever}} is violated. The larger is $\bar \varepsilon d$, the larger is the deviation considered. Finally, $F_{Y_{t}(0)|d_r=1}^{-1}$ represents the true population parameter, whereas `Bias' and `RMSE' stand for average (simulated) bias, and root-mean-squared, respectively.}}
\end{table}  

The results are shown in \hyperref[montecarlo3]{Table~\ref*{montecarlo3}} and I consider $\bar \varepsilon\in\{0.00,0.05,0.10,0.50\}$. For brevity, I report results only for the $F_{Y_{2}(0)|d_2=1, \rho=0}^{-1}(\tau)$, though all other results are consistent with those in \hyperref[montecarlo3]{Table~\ref*{montecarlo3}} and available upon request.

The results show that small violations of \hyperref[as:CPTnever]{Assumption~\ref*{as:CPTnever}} lead to only minor increases in bias and RMSE. For instance, when focusing on the $50$th quantile and considering the conditional parameter, the bias increases in magnitude from $0.012$ (for $\bar \varepsilon=0.00$) to $0.037$ when $\bar \varepsilon=0.05$ and to $0.086$ when $\bar \varepsilon=0.10$. The same applies to the RMSE. In contrast, for $\bar \varepsilon=0.50$ (a large deviation from \hyperref[as:CPTnever]{Assumption~\ref*{as:CPTnever}}), the bias increases substantially. Similar results are observed for the $25$th and $75$th quantiles.

Furthermore, as demonstrated in the previous subsection, disregarding covariates significantly increases the estimator's bias and RMSE, except in cases where the violation of \hyperref[as:CPTnever]{Assumption~\ref*{as:CPTnever}} is large. The poorer performance of the proposed estimator when conditioning on covariates in the presence of a large deviation from the Distributional PT assumption can be attributed to the substantial deviation in linear trends, which is not adequately captured by a single covariate. Since this covariate, by construction, has the same effect over time, it is unrelated to the trend in  $Y(0)$. Therefore, conditioning on $X$ increases estimation noise within a given quantile. \\

\noindent \textit{DGP 5}: In this last DGP, I assess the performance of the proposed estimator in scenarios where \hyperref[as:CPTnever]{Assumption~\ref*{as:CPTnever}} still holds, but now I examine the effects of small deviations from unconditional \hyperref[as:Copula]{Assumption~\ref*{as:Copula}}, similar to DGP 2 in \cite{Callawayetal2018}. Specifically, I assume that the untreated potential outcome follows the TWFE regression with covariates in \hyperref[DGP2]{(\ref*{DGP2})} and that:

\begin{equation*}
    \left(\eta_i, u_{i,t}, u_{i,r-1}\right)|d_r=1\sim N\left(\pmb{\mu}_r, V_r\right)
\end{equation*}

where $\pmb{\mu}_r=\left[r,0, 0\right]^T$ and 

$$V_r=\begin{pmatrix}
    1 & \rho_{r,u_t} & \rho_{r,u_{r-1}}  \\
    \rho_{r,u_t} & 1 & \rho_{u_t,u_{r-1}} \\
    \rho_{r,u_{r-1}} & \rho_{u_t,u_{r-1}} & 1 \\
\end{pmatrix} \quad r\in\{2,3, 4\}, t\in\{1,\dots,4\}$$

For simplicity, I assume $V_r$ is symmetric; for example, $cov(u_t,u_{r-1})=cov(u_{r-1},u_t)$. Consequently, $\left(\Delta_{r-1,t}y\left(0\right), Y_{r-1}\right)$ follows a bivariate normal distribution with correlation parameter given by $\rho_{u_t,u_{r-1}}+\rho_{r,u_t}-\rho_{r,u_{r-1}}-2$.

As noted by \cite{Callawayetal2018}, when the bivariate distribution is normal, then the copula is Gaussian and the dependence parameter corresponds to the correlation coefficient.

To evaluate the estimator's performance under deviations from the Copula Invariance Assumption, I set $\rho_{u_t,u_{r-1}}=0.5$ for all treated and never-treated units and define:

\begin{equation*}
   \rho_{r,u_t}= \begin{cases}
        \bar \rho d & \forall t\geq r \\
        0 & \forall t < r \\
    \end{cases}
\end{equation*}

where $d=0$ for never-treated units and $d=1$ for treated units (regardless of cohort), and $\bar \rho$ represents the degree of violation of Copula Invariance Assumption allowed. When $\bar \rho=0$, the assumption holds, while $\bar \rho \neq 0$ indicates a violation. The larger $\bar \rho$ is, the greater the deviation considered.   

\begin{table}[htbp]
\caption{\footnotesize Monte Carlo Results, DGP 5, for $F_{Y_{2}(0)|d_2=1, \rho=0}^{-1}(\tau)$ -- Violation Copula Invariance.}
\label{montecarlo4}
\resizebox{\textwidth}{!}{
\begin{tabular}{l*{10}{c}}
\toprule
\small
\vspace*{0.0002cm}	\\																								
\vspace*{0.0002cm}	\\																								
				&	& \multicolumn{3}{c}{\textbf{0.25}} 					&\multicolumn{3}{c}{\textbf{0.50}} 				& \multicolumn{3}{c}{\textbf{0.75}} 			\\							
	\cmidrule(rl){3-5} \cmidrule(rl){6-8} \cmidrule(rl){9-11} 																								
\vspace*{0.0025cm}	\\																								
	$\bar \rho$	& 	&	$F_{Y_{t}(0)|d_r=1}^{-1}$ 	&	Bias	&	Root MSE 	&		$F_{Y_{t}(0)|d_r=1}^{-1}$ 	&	Bias	&	Root MSE 	&		$F_{Y_{t}(0)|d_r=1}^{-1}$ 	&	Bias	&	Root MSE 	\\
  \vspace*{0.0000001cm}	\\																								
\cline{1-11}																									
  \vspace*{0.0025cm}	\\																								
	
\multirow{2}{4em}{\textbf{0.00}}	&	\textbf{Unc}	&	2.832	&	0.270	&	0.324	&	4.000	&	0.263	&	0.314	&	5.168	&	0.261	&	0.317  \\
                                    &	\textbf{Cond}	&	2.832	&	0.023	&	0.153	&	4.000	&	0.017	&	0.146	&	5.168	&	0.015	&	0.16	\\
																					
\multirow{2}{4em}{\textbf{0.05}}	&	\textbf{Unc}	&	2.812	&	0.290	&	0.341	&	4.000	&	0.263	&	0.314	&	5.188	&	0.241	&	0.302	\\
                                    &	\textbf{Cond}	&   2.812	&	0.042	&	0.157	&	4.000	&	0.017	&	0.146	&	5.188	&	-0.005	&	0.159	\\
																					
\multirow{2}{4em}{\textbf{0.10}}   &	\textbf{Unc}	&	2.793	&	0.308	&	0.357	&	4.000	&	0.263	&	0.314	&	5.207	&	0.222	&	0.287	\\
                                   &	\textbf{Cond}	 &   2.793	&	0.061	&	0.163	 &	 4.000	 &	 0.016	 &	0.146	&	5.207	&	-0.024	&	0.161	\\
																					
\multirow{2}{4em}{\textbf{0.50}}   &	\textbf{Unc}	&	2.651	&	0.449	&	0.484	&	4.000	&	0.263	&	0.314	&	5.349	&	0.080	&	0.197	\\
                                   &	\textbf{Cond}	 &	2.651	&	0.203	&	0.254	&	4.000	&	0.017	&	0.147	&	5.349	&	-0.166	&	0.23	\\
\bottomrule
\end{tabular}}
\floatfoot{\footnotesize{\textbf{Notes}: This table reports the Monte Carlo results of the DGP in \hyperref[DGP2]{Eq. (\ref*{DGP2})} for the estimator of the parameter $F_{Y_{2}(0)|d_2=1, \rho=0}^{-1}(\tau)$ in the setup with $T=R=4$ and violations of the Copula Invariance Assumption. Results are reported for the quantiles $.25$, $.50$, $.75$. Each Monte Carlo simulation uses $2,000$ bootstrap replications. Rows labeled `UNC' use the estimator based on the unconditional distributional PT assumption that ignores covariates. Here, $\bar \rho$ controls whether the Copula Invariance is violated. When $\bar \rho=0$, \hyperref[as:Copula]{Assumption~\ref*{as:Copula}} is not violated. When $\bar \rho \neq 0$, \hyperref[as:Copula]{Assumption~\ref*{as:Copula}} is violated. The larger is $\bar \varepsilon d$, the larger is the deviation considered. Finally, $F_{Y_{t}(0)|d_r=1}^{-1}$ represents the true population parameter, whereas `Bias' and `RMSE' stand for average (simulated) bias, and root-mean-squared, respectively.}}
\end{table}  

\hyperref[montecarlo4]{Table~\ref*{montecarlo4}} presents the results for $\bar \rho\in\{0.00,0.05,0.10,0.50\}$. For brevity, I report results only for $F_{Y_{2}(0)|d_2=1, \rho=0}^{-1}(\tau)$, though all other results align with those in \hyperref[montecarlo4]{Table~\ref*{montecarlo4}} and are available upon request.

Small violations of the Copula Invariance Assumption lead to minimal increases in the bias and RMSE is small. However, substantial violations (e.g., $\bar \rho=0.50$) cause a pronounced increase in bias.  For instance, considering the conditional parameter for the $0.25$th quantile, the bias rises from $0.061$ for $\bar\rho=0.10$ to $0.203$ for $\bar\rho=0.50$. Similar patterns emerge for the $0.75$th.

Conversely, the results for the $.50$th quantile are insensitive to deviations from the Copula Invariance Assumption. When allowing deviations from the Copula Invariance Assumption, only the variance of $Y_{it}(0)$ is affected. However, $\Phi^{-1}\left(0.50\right)=0$, thus we obtain $F_{Y_{t}(0)|d_r=1}^{-1}(\tau)= \left(\alpha_t +r\right)$.  These results align with those in \cite{Callawayetal2018}.

Lastly, in nearly all the cases considered, ignoring covariates significantly increases estimator bias and RMSE. The only exception is for the $0.75$th when $\bar\rho=0.50$, where the bias is lower under the unconditional parameter--a puzzling result.

%% file: Conclusion.tex
\section{Discussion}\label{Discussion_2}

The proposed method is only one of several possible ways to construct the counterfactual distribution. Estimating the counterfactual outcome is often challenging, as different estimators rely on different identifying assumptions.\footnote{See \cite{Imbens2009} for a review of common assumptions in the causal inference literature for estimating counterfactual outcomes.} The performance of an estimator can vary depending on the context, so researchers should assess whether the proposed method is suitable for addressing their specific research question.  

Several existing approaches identify the untreated potential outcome distribution for the treated group while allowing unobserved characteristics to vary across treated units. Using the intuition presented in this paper, these methods can be extended to the staggered treatment adoption setting.

The Changes-in-changes (CIC) model proposed by \cite{AtheyImb2006} relates the outcome without intervention to the individual's group, time, and unobservable characteristics through a monotonic production function. While accounting for the selection on unobservables, the CIC model assumes that unobservable traits' distribution within a group is stable over time. \cite{BonhommeSauder}, on the other hand, propose an estimator of the distributional treatment effects by requiring that the production function mapping groups, time, and covariates into outcome is additive.

Unlike \cite{AtheyImb2006} and \cite{BonhommeSauder}, the approach proposed in this paper does not restrict the functional form that relates groups, time, and covariates. Moreover, unlike \cite{BonhommeSauder}, the proposed method does not require the unobservable time-varying component to be independent of treatment assignment. Nor does it assume that the time-varying unobservables are independent of time-invariant unobservables, conditional on treatment assignment. 

The presented estimator requires only that the evolution of the untreated potential outcome be (conditionally) independent of treatment assignment while still allowing for serial correlation in the error term. Additionally, it permits the time-varying shock to be correlated with the individual (unobserved) heterogeneity. In this sense, the estimator proposed is more general than the other existing approaches, though this generality comes at the cost of an additional assumption regarding the missing dependence (copula) between the change in the untreated potential outcome and its pre-treatment level. 

On the other hand, unlike the CIC model, the proposed approach is not scale-invariant. Additionally, in contrast to \cite{BonhommeSauder}, the method does not allow for returns to unobserved skills to vary after introducing the policy.

It is also worth emphasizing that this estimator can complement existing methods for estimating the ATT when analyzing the impact of policies introduced in a staggered fashion. Given its similarity to \cite{Callaway2021}, I compare my estimator primarily to theirs, as the proposed approach extends their methodology.\footnote{Interested readers may refer to \cite{Wooldridge2021} for a comparison of the methods proposed by \cite{Callaway2021}, \cite{Sun2021}, and \cite{Wooldridge2021}.}

As discussed in previous sections, the proposed estimator complements existing estimators when heterogeneity along the outcome distribution is expected. While ATT is a widely used measure for evaluating policy effects, focusing on distributional treatment effects on the treated allows for broader applications, such as in equity assessments of policy interventions, making it relevant for policymakers. For example, in studying the effects of minimum wage policies on teen employment, as in \cite{Callaway2021}, the proposed estimator may offer deeper insights by capturing effects beyond the mean--particularly on the lower deciles of the income distribution.

Although the proposed estimator offers advantages in settings where treatment effects are heterogeneous across the outcome distribution, it requires stronger identifying assumptions for consistency. Although Assumptions \hyperref[as:irreversibility]{\ref*{as:irreversibility}}-\hyperref[as:NA]{~\ref*{as:NA}} closely align with those in \cite{Callaway2021}, additional restrictions are needed beyond the Conditional Mean PT assumption to achieve point identification of $F_{y_t(0)|d_r=1}$. In particular, the estimator proposed is sensitive to strong violations of the Copula Invariance assumption \citep[not required in][]{Callaway2021} and may underperform, compared to the method proposed by \cite{Callaway2021}, when the PT assumption holds on average but not across the entire distribution.

Overall, when evaluating the effect of policy interventions, researchers should assess the plausibility of the identifying assumptions underlying this method relative to other available approaches.

\section{Conclusion}\label{Conclusion_2}

In this paper, I provide a method to recover the entire distribution of the untreated potential outcome for the treated group in nonexperimental settings with staggered treatment adoption. To do so, I build on the idea behind the group-time average treatment effect estimator proposed by \cite{Callaway2021} and generalize the existing quantile treatment effects on the treated estimator proposed by \cite{CallawayLi2019}. 

Once the QTT parameters of interest are constructed, they can be aggregated using the same schemes suggested by \cite{Callaway2021} to highlight heterogeneity along specific dimensions--such as how treatment effects vary with the length of exposure to treatment. 

Further, I show that, beyond the QTT, alternative approaches that anonymously summarize the quantiles of the outcome distribution can be developed within this framework once the entire counterfactual distribution is identified. In particular, I consider tests for stochastic dominance rankings, which--unlike the QTT--do not rely on the assumption of rank invariance. In doing so, this paper bridges the literature on causal inference with that on inequality measurement. Depending on the parameter of interest, I advocate that researchers employ the most suitable testing procedures to perform statistical inference.

Identification is achieved by extending the commonly used parallel trends assumption to the entire distribution of the untreated potential outcome. However, this extended assumption alone is not sufficient for point identification \citep{Fan2012}. Therefore, an additional assumption is required regarding the missing dependence structure (or copula) between the change in the untreated potential outcome and its pretreatment level. To restore point identification, I follow \cite{Callawayetal2018} and assume that this missing dependence is (conditionally) independent of treatment assignment.

I show that, under the assumptions presented in this paper, the proposed method's performance is poor when the sample size is small, but this improves substantially as $n$ increases. Furthermore, the proposed method performance remains almost unaffected even in cases with minor deviations from the main identifying assumptions. These findings align with the theoretical predictions established by \cite{Callawayetal2018, CallawayLi2019, Callaway2021}.

%% file: AppendixA.tex
\section*{Appendix A -- Proofs}\label{AppendixA_2}
\renewcommand{\theequation}{A.\arabic{equation}}
\setcounter{equation}{7} 

\subsection*{Identification}
\subsubsection*{Identification without covariates}
In this subsection, I will prove the result obtained in \hyperref[Th1]{Theorem \ref*{Th1}}. This proof is adapted from \cite{CallawayLi2019}. To prove this theorem, I will use two results of Sklar's Theorem: Lemma A.1. and Lemma A.2. in Appendix A in \cite{CallawayLi2019}. To save space, I refer the reader to their paper.

\begin{proof}[Proof of Theorem 1]
For notational convenience, I will assume that there is no treatment anticipation (i.e., $\rho=0$) and abbreviate the joint pdf at time $t$ of the change in untreated potential outcome and the pre-treatment untreated potential outcome for treated group $r$ as $f_{t| d_r=1}(\cdot,\cdot)=f_{Y_t\left(0\right)-Y_{r-\rho-1}\left(0\right),Y_{r-\rho-1}\left(0\right)|d_r=1}$. Similarly, this will be, instead, the same pdf for the never-treated $f_{t|C=1}(\cdot,\cdot)=f_{Y_t\left(0\right)-Y_{r-\rho-1}\left(0\right),Y_{r-\rho-1}\left(0\right)|C=1}$. Further, I will denote the copula pdfs between the change in untreated potential outcome and the pre-treatment untreated potential outcome as $c_{t| d_{r}=1}(\cdot,\cdot)=c_{Y_t\left(0\right)-Y_{r-\rho-1}\left(0\right),Y_{r-\rho-1}\left(0\right)|d_{r}=1}$ and $c_{t| C=1}(\cdot,\cdot)=c_{Y_t\left(0\right)-Y_{r-\rho-1}\left(0\right),Y_{r-\rho-1}\left(0\right)|C=1}$. Assuming also that $\Delta_{\left[r-1,t\right]}Y(0)$ has support in $\Delta\mathcal{Y}$ and $Y_{t-1}(0)$ in $\mathcal{Y}$, then:

\begin{equation*}
\begin{aligned}
F_{Y_t(0) |d_r=1} & = \P\left(Y_t(0) \leq y |d_r=1\right) \\
& = \P\left(\Delta_{\left[r-1,t\right]}Y(0)+Y_{r-1}(0)\leq y|d_r=1\right) \\
& = \mathbb{E}\left[ \mathbbm{1}\left\{\Delta_{\left[r-1,t\right]}Y(0) \leq y-Y_{r-1}(0) |d_r=1\right\}\right] \\
& =\int_{\mathcal{Y}} \int_{\Delta \mathcal{Y}}  \mathbbm{1}\left\{\delta \leq y-y^{\prime}\right\} f_{t| d_{r}=1}\left(\delta, y^{\prime}\right) d \delta d y^{\prime} \\
& =\int_{\mathcal{Y}} \int_{\Delta \mathcal{Y}}  \mathbbm{1}\left\{\delta \leq y-y^{\prime}\right\} c_{t| d_{r}=1}\left(F_{\Delta_{\left[r-1,t\right]}Y(0)|d_r=1}, F_{Y_{r-1}(0)|d_r=1}\right) \\
&\quad{}\times f_{\Delta_{\left[r-1,t\right]}Y(0)|d_r=1}(\delta)f_{Y_{r-1}(0)|d_r=1}(y^{\prime}) d \delta d y^{\prime} \quad{} \quad{} (A.1) \\
& =\int_{\mathcal{Y}} \int_{\Delta \mathcal{Y}}  \mathbbm{1}\left\{\delta \leq y-y^{\prime}\right\} c_{t| C=1}\left(F_{\Delta_{\left[r-1,t\right]}Y(0)|d_r=1}, F_{Y_{r-1}(0)|d_r=1}\right) \\
&\quad{}\times f_{\Delta_{\left[r-1,t\right]}Y(0)|d_r=1}(\delta)f_{Y_{r-1}(0)|d_r=1}(y^{\prime}) d \delta d y^{\prime} \quad{} \quad{} (A.2) \\
& =\int_{\mathcal{Y}} \int_{\Delta \mathcal{Y}}  \mathbbm{1}\left\{\delta \leq y-y^{\prime}\right\} f_{t| C=1}(F^{-1}_{\Delta_{\left[r-1,t\right]}Y(0)|C=1}\left(F_{\Delta_{\left[r-1,t\right]}Y(0)|d_r=1}(\delta)\right),  \\
& \quad{} \quad{} F^{-1}_{Y_{r-1}(0)|C=1}\left(F_{Y_{r-1}(0)|d_r=1} (y^{\prime})\right)) \\
&\quad{}\times \frac{f_{\Delta_{\left[r-1,t\right]}Y(0)|d_r=1}(\delta)}{f_{\Delta_{\left[r-1,t\right]}Y(0)|C=1}\left(F_{\Delta_{\left[r-1,t\right]}Y(0)|d_r=1}(\delta)\right)} \\
& \quad{} \times \frac{f_{Y_{r-1}(0)|d_r=1}(y^{\prime}) }{f_{Y_{r-1}(0)|C=1}\left(F^{-1}_{Y_{r-1}(0)|C=1}\left(F_{Y_{r-1}(0)|d_r=1} (y^{\prime})\right))\right) } d \delta d y^{\prime} \quad{} \quad{} (A.3) \\
\end{aligned}
\end{equation*}

Equation (A.1) exploits Lemma A.1 in \cite{CallawayLi2019} to write the joint distribution using a copula pdf. Equation (A.2) exploits \hyperref[as:Copula]{Assumption \ref*{as:Copula}} to retrieve the missing dependence between the change in untreated potential outcome and the pre-treatment value of untreated potential outcome for treated group $r$. Lastly, equation (A.3) employs Lemma A.2 in \cite{CallawayLi2019} to rewrite the copula pdf as the joint distribution for the never-treated group.

Let us now make the following change of variables to simplify computations. Specifically, let us denote with: 

\begin{equation*}
    u=F^{-1}_{\Delta_{\left[r-1,t\right]}Y(0)|C=1}\left(F_{\Delta_{\left[r-1,t\right]}Y(0)|d_r=1}(\delta)\right), \quad v= F^{-1}_{Y_{r-1}(0)|C=1}\left(F_{Y_{r-1}(0)|d_r=1} (y^{\prime})\right)
\end{equation*}

The above notation then implies the following equalities:
\begin{enumerate}
    \item $\delta=F^{-1}_{\Delta_{\left[r-1,t\right]}Y(0)|d_r=1}\left(F_{\Delta_{\left[r-1,t\right]}Y(0)|C=1}(u)\right)$
    \item $y^{\prime}= F^{-1}_{Y_{r-1}(0)|d_r=1}\left(F_{Y_{r-1}(0)|C=1} (v)\right)$
     \item $\frac{d\delta}{du}=\frac{f_{\Delta_{\left[r-1,t\right]}Y(0)|C=1}(u)}{f_{\Delta_{\left[r-1,t\right]}Y(0)|d_{r}=1}\left(F^{-1}_{\Delta_{\left[r-1,t\right]}Y(0)|d_r=1}\left(F_{\Delta_{\left[r-1,t\right]}Y(0)|C=1}(u)\right)\right)}$
    \item $\frac{d y^{\prime}}{dv}= \frac{f_{Y_{r-1}(0)|C=1}(v)}{f_{Y_{r-1}(0)|d_{r}=1}\left(F^{-1}_{Y_{r-1}(0)|d_r=1}\left(F_{Y_{r-1}(0)|C=1} (v)\right)\right)}$
\end{enumerate}

If we plug (1)-(4) in equation (A.3), we obtain the following equalities:
\begin{equation*}
\begin{aligned}
   &  =\int_{\mathcal{Y}} \int_{\Delta \mathcal{Y}}  \mathbbm{1}(F^{-1}_{\Delta_{\left[r-1,t\right]}Y(0)|d_r=1}\left(F_{\Delta_{\left[r-1,t\right]}Y(0)|C=1}(u)\right)\leq y- \\
  & \quad F^{-1}_{Y_{r-1}(0)|d_r=1}\left(F_{Y_{r-1}(0)|C=1} (v)\right)) \times f_{t|C=1}(u,v) du dv \quad (A.4) \\
  & = \E \left[ \mathbbm{1} \left(F^{-1}_{\Delta_{\left[r-1,t\right]}Y(0)|d_r=1}\left(F_{\Delta_{\left[r-1,t\right]}Y(0)|C=1}\left(\Delta_{\left[r-1,t\right]}Y(0)\right)\right)\right. \leq y -F^{-1}_{Y_{r-1}\left(0 \right) |d_{r}=1} \times \right.\\
            & \quad \left. \left.  \left(F_{Y_{r-1}\left(0\right) |C=1} \left(Y_{r-1}(0)\right)\right)\right]  |C=1 \right] \quad (A.5) \\
            & = \E \left[ \mathbbm{1} \left(F^{-1}_{\Delta_{\left[r-1,t\right]}Y(0)|C=1}\left(F_{\Delta_{\left[r-1,t\right]}Y(0)|C=1}\left(\Delta_{\left[r-1,t\right]}Y(0)\right)\right)\right. \leq y -F^{-1}_{Y_{r-1}\left(0 \right) |d_{r}=1} \times \right.\\
            & \quad \left. \left.  \left(F_{Y_{r-1}\left(0\right) |C=1} \left(Y_{r-1}(0)\right)\right)\right]  |C=1 \right] \quad (A.6) \\        
             & = \E \left[ \mathbbm{1} \left(\Delta_{\left[r-1,t\right]}Y(0)\right) \leq y -F^{-1}_{Y_{r-1}\left(0 \right) |d_{r}=1}  \left(F_{Y_{r-1}\left(0\right) |C=1} \left(Y_{r-1}(0)\right)\right)  |C=1 \right] \quad (A.7)
\end{aligned}
\end{equation*}
Where equation (A.4) comes from substituting previous identities in equation (A.3); equation (A.5) comes from the definition of $\E(\cdot)$; equation (A.6) replaces the unknown distribution $F^{-1}_{\Delta_{\left[r-1,t\right]}Y(0)|d_r=1}$ with the distribution of the change in untreated potential outcome for the never-treated group thanks to \hyperref[as:CPTnever]{Assumption \ref*{as:CPTnever}} holding unconditionally, and lastly equation (A.7) proves the result as each of these distributions of untreated potential outcomes is identified by their sample counterparts. This proves identification of $ F_{Y_t(0) |d_r=1}$.\\
\end{proof}

\begin{proof}[Alternative Proof of Theorem 1]
An alternative and more direct approach to proving the identification of $ F_{Y_t(0) |d_r=1}$ is to exploit \hyperref[as:Continuity]{Assumption \ref*{as:Continuity}} and Sklar's Theorem, as done in \cite{Callawayetal2018}.

To see this, recall that for every $y\in supp\left(Y_t(0)|d_r=1\right)$, we can express $F_{Y_t(0) |d_r=1}$:

\begin{equation}
    \begin{aligned}\label{A8}
F_{Y_t(0) |d_r=1} & = \P\left(Y_{it}(0) \leq y |d_r=1\right) \\
& = \P\left(\Delta_{\left[r-1,t\right]}Y_i(0)+Y_{i,r-1}(0)\leq y|d_r=1\right) \\
\end{aligned}
\end{equation}

Under the continuity assumption, we can write, for units treated in cohort $r$:
\begin{equation}\label{A9}
    \Delta_{\left[r-1,t\right]}Y_i(0)=F^{-1}_{\Delta_{\left[r-1,t\right]}Y(0)| d_r=1}\left(u_i^r\right) \quad \text{and} \quad Y_{i,r-1}(0)=F^{-1}_{Y_{r-1}(0)| d_r=1}\left(v_i^r\right)
\end{equation}
where $u_i^r \equiv F_{\Delta_{\left[r-1,t\right]}Y(0)| d_r=1}\left(\Delta_{\left[r-1,t\right]}Y_i(0)\right)$ and $v_i^r \equiv F_{Y_{i,r-1}(0)| d_r=1}\left(Y_{i,r-1}(0)| d_r=1\right)$.

Similarly, for the never-treated:
\begin{equation}\label{A10}
    \Delta_{\left[r-1,t\right]}Y_i(0)=F^{-1}_{\Delta_{\left[r-1,t\right]}Y(0)| C=1}\left(u_i^0\right) \quad \text{and} \quad Y_{i,r-1}(0)=F^{-1}_{Y_{r-1}(0)| C=1}\left(v_i^0\right)
\end{equation}
where $u_i^0 \equiv F_{\Delta_{\left[r-1,t\right]}Y(0)| C=1}\left(\Delta_{\left[r-1,t\right]}Y_i(0)\right)$ and $v_i^0 \equiv F_{Y_{i,r-1}(0)| C=1}\left(Y_{i,r-1}(0)| C=1\right)$.

Thus, substituting these into \hyperref[A8]{(\ref*{A8})}, we obtain:

\begin{equation}\label{A11}
F_{Y_t(0) |d_r=1}  = \P\left(F^{-1}_{\Delta_{\left[r-1,t\right]}Y(0)| d_r=1}\left(u_i^r\right)+F^{-1}_{Y_{r-1}(0)| d_r=1}\left(v_i^r\right)\leq y|d_r=1\right) 
\end{equation}

For $d_r=1$, the joint distribution $F_{\left(u_i^r, v_i^r\right)|d_r=1}$ is unknown. However, by Sklar's Theorem, we express it in terms of conditional copula  $C_{\Delta  Y_{t}(0), Y_{t-1}(0)| X, d_r=1}$. By the unconditional version of assumption \hyperref[as:Copula]{Assumption \ref*{as:Copula}}, we replace $C_{\Delta  Y_{t}(0), Y_{t-1}(0)| d_r=1}$ with $C_{\Delta  Y_{t}(0), Y_{t-1}(0)| C=1}$. 

Substituting this into \hyperref[A11]{(\ref*{A11})}, we obtain:

\begin{equation*}
F_{Y_t(0) |d_r=1}  = \P\left(F^{-1}_{\Delta_{\left[r-1,t\right]}Y(0)| d_r=1}\left(u_i^0\right)+F^{-1}_{Y_{r-1}(0)| d_r=1}\left(v_i^0\right)\leq y|C=1\right) 
\end{equation*}

While $F^{-1}_{Y_{r-1}(0)| d_r=1}$ is observable from the data, $F^{-1}_{\Delta_{\left[r-1,t\right]}Y_{(0)}| d_r=1}$ is not. Using \hyperref[as:CPTnever]{Assumption \ref*{as:CPTnever}}, we replace the unknown distribution $F^{-1}_{\Delta_{\left[r-1,t\right]}Y(0)|d_r=1}$ with the distribution of the change in untreated potential outcome for the never-treated group. Further, using \hyperref[A10]{(\ref*{A10})} and the definitions of $u_i^0$ and $v_i^0$, we obtain:

\begin{equation*}
             F_{ Y_{t}\left( 0 \right)|d_r=1} = \P  \left(\Delta_{\left[r-1,t\right]}Y_i(0)  + F^{-1}_{Y_{r-1}\left(0 \right) |d_{r}=1} \left(F_{Y_{r-1}\left(0\right) |C=1} \left(Y_{i,r-1}(0)\right)\right) \leq y |C=1 \right)
         \end{equation*}
which completes the proof.
\end{proof}

\subsubsection*{Identification with covariates}
In the first part of this subsection I will prove the identification result shown \hyperref[Prop1]{Proposition \ref*{Prop1}}; in the second part of this subsection, instead, I will prove \hyperref[Prop2]{Proposition \ref*{Prop2}}. Also in these cases, proofs are adapted from \cite{CallawayLi2019}.

\begin{proof}[Proof of Proposition 1]
    As pointed out in \hyperref[Identification]{Section \ref*{Identification}}, all the results obtained in \hyperref[Th1]{Theorem \ref*{Th1}} are still valid. The only part that changes is equation (A.6), which used an unconditional version of \hyperref[as:CPTnever]{Assumption \ref*{as:CPTnever}} to reach identification of $F_{\Delta Y(0) |d_r=1 }$. Now, this object is identified by the reweighted distribution in \hyperref[propscoreFirpo]{Eq. (\ref*{propscoreFirpo})}. To show that the results obtained in \hyperref[Th1]{Theorem \ref*{Th1}} are still valid what we need to prove is that $F_{\Delta Y(0) |d_r=1 }=F^p_{\Delta Y(0) |d_r=1 }$. To prove it, let us exploit the definition of $F_{\Delta Y(0) |d_r=1 }$:
    \begin{equation*}
        \begin{aligned}
            F_{\Delta Y(0) |d_r=1 } & = \P(\Delta Y_t(0) \leq  \delta |d_r=1 ) \\
            & =\frac{\P(\Delta Y_t(0) \leq  \delta,  d_r=1 )}{p_r} \quad (A.12) \\
            & = \E\left(\frac{\P(\Delta Y_t(0) \leq  \delta , d_r=1| X )}{p_r} \right) \\
            & = \E\left(\frac{P_r(X)}{p_r} \P(\Delta Y_t(0) \leq  \delta |  d_r=1, X ) \right) \\
            & = \E\left(\frac{P_r(X)}{p_r} \P(\Delta Y_t(0) \leq  \delta |  X, C=1 ) \right) \quad (A.13) \\
            & = \E\left(\frac{P_r(X)}{p_r} \E\left[ C \mathbbm{1}\{\Delta Y \leq  \delta\} |  X, C=1 \right]  \right) \quad (A.14) \\
            & = \E\left(\frac{P_r(X)}{p_r(1-p_r(X))} \E\left[ C \mathbbm{1}\{\Delta Y_t \leq  \delta\}|  X\right]  \right)\\
             & = \E\left(\frac{ C p_r(X)}{p_r(1-p_r(X))}  \mathbbm{1}\{\Delta Y_t \leq  \delta\}  \right) \quad (A.11)
        \end{aligned}
    \end{equation*}
    
where $p_r=p(d_r=1)$ denotes the probability of being first treated in cohort $r$, and $p_{r}(X)=\P\left(d_r=1|X, d_r+C=1\right)$ denotes the generalized propensity score, as defined in \hyperref[as:Overlap]{Assumption \ref*{as:Overlap}}. Specifically, while $p_r$ captures the fraction of treated units in a given cohort, $p_{r}(X)$ models the conditional probability of belonging to cohort $r$ and either being part of cohort $r$ or the never-treated group.

In Equation (A.12) I exploited the definition of conditional probability; Equation (A.13) holds since \hyperref[as:CPTnever]{Assumption \ref*{as:CPTnever}} holds;  in  Equation (A.14), I exploited the definition of probability, and then I multiplied by $C$ (this holds since $\E(\cdot)$ is conditionally on $C=1$). Further, by conditioning on  $C=1$, we can rewrite the potential outcome as the observed outcome. The last equality exploits the Law of Iterated Expectations and concludes the proofs.   \\
\end{proof}

\begin{proof}[Proof of Proposition 2]
    The proof of Proposition 2 follows directly from \hyperref[Th1]{Theorem \ref*{Th1}}, where now all the steps hold after conditioning on covariates.  \\
\end{proof}

%% file: Appendix_repeated.tex
\clearpage
\newpage

\section*{Appendix B -- Additional Results for Repeated Cross Sections}\label{Appendix_rep_cross}
\renewcommand{\theassumption}{B.\arabic{assumption}}
\setcounter{assumption}{0}

In this section, I extend the identification results of this paper to the setting where repeated cross sections are available. To this end, I will follow \cite{Callawayetal2018} and \cite{Callaway2021}.

Specifically, I assume that for each cross-sectional unit, the researcher has access to $\left(Y, d_r,\dots, d_T, C, t, X\right)$ where $r=q,\dots,T$ and $t=1,\dots,T$ is the  period in which unit $i$ is observed. Additionally, if we denote with $S_t$ a dummy variable taking value $1$ if an observation is observed in period $t$ and zero otherwise.

Now, I assume that a random sample is available in each period.

\begin{assumption}\label{as:RS_cross}
Conditional on period $t$, $\left(Y, d_r,\dots, d_T, C, X\right)$ are cross sectionally independent and identically distributed for all $t\in\{1,\dots,T\}$, where $\left(d_r,\dots, d_T, C, X\right)$ is invariant to $t$.
\end{assumption}

\hyperref[as:RS_cross]{Assumption \ref*{as:RS_cross}} is identical to that done in \cite{Callaway2021} and means that the pooled cross section is composed of independent draws from the following mixture distribution:

\begin{equation*}
    F_M(Y, d_r,\dots, d_T, c, t, x)=\sum_{t=1}^T \P(S_t=1) F_{Y, d_r,\dots, d_T, C, t, X}\left(Y, d_r,\dots, d_T, c, x | t\right)
\end{equation*}

This assumption is also similar to the assumption in \cite{Abadie2005} and \cite{Santanna2020}, and excludes the possibility of compositional changes over time.

For simplicity, suppose that covariates do not play a role in identification. Further, assume without loss of generality that there is no treatment anticipation (i.e., $\rho=0$). To extend the main identification result from \hyperref[Th1]{Theorem \ref*{Th1}}, we impose an additional requirement on the data-generating process. Even if Assumptions \hyperref[as:irreversibility]{\ref*{as:irreversibility}}, \hyperref[as:RS_cross]{\ref*{as:RS_cross}}, \hyperref[as:Continuity]{\ref*{as:Continuity}}, and an unconditional version of Assumptions \hyperref[as:NA]{\ref*{as:NA}}- \hyperref[as:Copula]{\ref*{as:Copula}} hold, the difference $Y_{it}-Y_{i,r-1}$ is not observed for the same unit in a pooled cross section. 

To achieve point identification, following \cite{Callawayetal2018}, I impose a \textit{rank invariance} assumption on potential outcomes over time. This assumption allows us to recover the (unobserved) outcome at period $t$ by exploiting the rank of the outcome observed in $r-1$. It assumes that units preserve their relative position in the distribution of $Y$ over time.

Although rank invariance is a strong assumption and is often rejected in empirical data (as discussed in the main text), it is necessary to recover the unknown distribution at time $t$.

\begin{corollary}\label{Corollary1}
\textit{Suppose we have access to repeated cross sectional data, specifically} $\{\left(Y,d_{ir}, C_i\right)\}_{i=1}^{n^s}$ \textit{for period} $s\in\{r-1,t\}$ \textit{where} $r=q,\dots,T$,  $t=1,\dots,T$, \textit{and} $n^s$ \textit{denotes the sample size of the cross section. Suppose further that Assumptions 1, 6, B1, and the unconditional version of Assumptions 3-5 hold.}

\textit{If the copula of} $\left(Y_{i,r-1}(0), Y_{i,t}(0)|C=1\right)$ \textit{satisfies rank invariance, then for every} $(u,v)\in\left[0,1\right]^2$  

    \begin{equation*}
    C_{Y_{i,r-1}(0), Y_{i,t}(0)|C=1}\left(u,v\right)=min\{u,v\}
    \end{equation*}
\textit{Thus, for} $y\in supp\left(Y_{i,t}(0)| d_r=1\right)$, we obtain
\begin{equation*}
    F_{Y_t(0)|d_r=1}(y)= \P \{\tilde{\Delta}_{[r-1,t]} Y(0)+ F^{-1}_{Y_{r-1}(0)|d_r=1}\left(F_{Y_{r-1(0)|C=1}}\left(Y_{r-1}(0)\right)\right) \leq y| C=1\}
\end{equation*}

\textit{where} $\tilde{\Delta}_{[r-1,t]} Y(0)\equiv F^{-1}_{ Y_{t}(0)|C=1}\left(F_{Y_{r-1(0)|C=1}}\left(Y_{r-1}(0)\right)\right)- Y_{r-1}$.
\end{corollary}
 
As noted in \cite{Callawayetal2018}, the rank invariance assumption on the copula of $\left(Y_{i,r-1}(0), Y_{i,t}(0)|C=1\right)$ neither implies nor is implied by \hyperref[as:Copula]{Assumption \ref*{as:Copula}}. 

\begin{proof}[Proof of Corollary 1]
Under Assumptions 1, 6, B1, and the unconditional version of Assumptions 3-5, the result from \hyperref[Th1]{Theorem \ref*{Th1}} holds:
    \begin{equation*}
        F_{ Y_{t}\left( 0 \right)|d_r=1}  = \P \{\Delta_{\left[r-1,t\right]}Y(0) + F^{-1}_{Y_{r-1}\left(0 \right) |d_{r}=1} \left(F_{Y_{r-1}\left(0\right) |C=1} \left(Y_{r-1}(0)\right)\right)\leq y |C=1 \}
         \end{equation*}
for $y\in supp\left(Y_{i,t}(0)| d_r=1\right)$.

However, we are working with repeated cross sections, $\Delta_{\left[r-1,t\right]}Y(0)$ cannot be identified from the observed outcome for the never-treated, as in \hyperref[Th1]{Theorem \ref*{Th1}}. To solve for this issue, we impose the rank invariance assumption, which yields:
\begin{equation*}
     F_{ Y_{t}\left( 0 \right)|C=1}\left( Y_{t}\left( 0 \right)\right)= F_{Y_{r-1}\left( 0 \right)|C=1}\left(Y_{r-1}\left( 0 \right)\right)
\end{equation*}

since both $F_{ Y_{t}\left( 0 \right)|C=1}\left(\cdot\right)$ and $F_{Y_{r-1}\left( 0 \right)|C=1}\left(\cdot\right)$ are identifiable from observed outcomes, we can express $\Delta_{\left[r-1,t\right]}Y(0)$ for units with $C=1$ as 

$$\tilde{\Delta}_{[r-1,t]} Y(0)\equiv F^{-1}_{ Y_{t}(0)|C=1}\left(F_{Y_{r-1(0)|C=1}}\left(Y_{r-1}(0)\right)\right)- Y_{r-1}$$

which completes the proof.
\end{proof}

This result extends naturally to settings that include covariates. Specifically, while the identification result in Proposition 2 can be generalized by ensuring that all steps hold conditional on $X$, extending Proposition 1 requires a modified definition of the generalized propensity score, as outlined in Appendix B of \cite{Callaway2021}.

%% file: AppendixB.tex
\clearpage
\section*{Appendix C -- Additional Simulation Results}\label{AppendixB_2}

\renewcommand{\thetable}{C.\arabic{table}}
\setcounter{table}{0}
\begin{table}[htbp]
\center
\small
\resizebox{0.9\textwidth}{!}{
\caption{Monte Carlo Results, DGP 1, $\tau=.5$. Setup with $T=R=4$.}
\label{MC_dgp1_appendix}
\begin{tabular}{l*{6}{c}}
\toprule
\vspace*{0.0002cm}	\\																								
		&	&	&		 \multicolumn{2}{c}{$n=100$} 				&	 \multicolumn{2}{c}{$n=1,000$} 			\\
			\cmidrule(rl){4-5}						\cmidrule(rl){6-7}				
\vspace*{0.0025cm}	\\												
		&	&	$F_{ Y_{t}(0)|d_r=1}^{-1}$ 	&	Bias	&	Root MSE 	&	Bias	&	Root MSE 	\\
	&	$F_{Y_{2}(0)|d_2=1}^{-1}$	&	4	&	0.097	&	0.489	&	0.007	&	0.15	\\
	&	$F_{Y_{3}(0)|d_2=1}^{-1}$	&	5	&	0.089	&	0.491	&	0.009	&	0.148	\\
	&	$F_{Y_{4}(0)|d_2=1}^{-1}$	&	6	&	0.097	&	0.492	&	0.008	&	0.148	\\
	&	$F_{Y_{3}(0)|d_3=1}^{-1}$	&	6	&	0.087	&	0.451	&	0.013	&	0.133	\\
	&	$F_{Y_{4}(0)|d_3=1}^{-1}$	&	7	&	0.095	&	0.451	&	0.012	&	0.133	\\
	&	$F_{Y_{4}(0)|d_4=1}^{-1}$	&	8	&	0.101	&	0.416	&	0.015	&	0.126	\\
\bottomrule
\end{tabular}
\floatfoot{\footnotesize{\textbf{Notes}: This table reports the Monte Carlo results of the DGP in \hyperref[DGP1]{Eq. (\ref*{DGP1})} in the setup with $T=R=4$ for the quantile $\tau=.5$. Each Monte Carlo simulation uses $2,000$ bootstrap replications. Rows labeled 'UNC' use the estimator based on the unconditional distributional PT assumption that ignores covariates. Finally, $F_{ Y_{t}(0)|d_r=1}^{-1}$ represents the true population parameter, whereas `Bias' and `RMSE' stand for average (simulated) bias, and root-mean-squared, respectively.}}}
\end{table}  

\begin{table}[!htbp]
\center
\small
\resizebox{0.8\textwidth}{!}{
\caption{Monte Carlo Results, DGP 2, $\tau=.5$. Setup with $T=R=4$.}
\label{MC_dgp2_appendix}
\begin{tabular}{l*{6}{c}}
\toprule
\vspace*{0.0002cm}	\\																								
	&	&	&		 \multicolumn{2}{c}{$n=100$} 				&	 \multicolumn{2}{c}{$n=1,000$} 			\\
			\cmidrule(rl){4-5}						\cmidrule(rl){6-7}				
\vspace*{0.0025cm}	\\												
		&	&	$F_{ Y_{t}(0)|d_r=1}^{-1}$ 	&	Bias	&	Root MSE 	&	Bias	&	Root MSE 	\\
\multirow{6}{4em}{\textbf{Unc}}	&	$F_{Y_{2}(0)|d_2=1}^{-1}$	&	4	&	0.341	&	0.684	&	0.25	&	0.304	\\
	&	$F_{Y_{3}(0)|d_2=1}^{-1}$	&	5	&	0.344	&	0.686	&	0.248	&	0.303	\\
	&	$F_{Y_{4}(0)|d_2=1}^{-1}$	&	6	&	0.341	&	0.689	&	0.248	&	0.302	\\
	&	$F_{Y_{3}(0)|d_3=1}^{-1}$	&	6	&	0.469	&	0.723	&	0.371	&	0.41	\\
	&	$F_{Y_{4}(0)|d_3=1}^{-1}$	&	7	&	0.47	&	0.728	&	0.372	&	0.41	\\
	&	$F_{Y_{4}(0)|d_4=1}^{-1}$	&	8	&	0.571	&	0.794	&	0.479	&	0.509	\\
 \vspace*{0.0025cm}	\\												
\multirow{6}{4em}{\textbf{Cond}}	&	$F_{Y_{2}(0)|d_2=1}^{-1}$	&	4	&	0.091	&	0.515	&	0.011	&	0.147	\\
	&	$F_{Y_{3}(0)|d_2=1}^{-1}$	&	5	&	0.099	&	0.522	&	0.007	&	0.147	\\
	&	$F_{Y_{4}(0)|d_2=1}^{-1}$	&	6	&	0.091	&	0.519	&	0.005	&	0.145	\\
	&	$F_{Y_{3}(0)|d_3=1}^{-1}$	&	6	&	0.099	&	0.508	&	0.014	&	0.156	\\
	&	$F_{Y_{4}(0)|d_3=1}^{-1}$	&	7	&	0.098	&	0.515	&	0.012	&	0.151	\\
	&	$F_{Y_{4}(0)|d_4=1}^{-1}$	&	8	&	0.09	&	0.507	&	0.007	&	0.153	\\
\bottomrule
\end{tabular}}
\floatfoot{\footnotesize{\textbf{Notes}: This table reports the Monte Carlo results for all the parameters of the DGP in \hyperref[DGP2]{Eq. (\ref*{DGP2})} in the setup with $T=R=4$ for the quantile $\tau=.5$. Each Monte Carlo simulation uses $2,000$ bootstrap replications. Rows labeled `UNC' use the estimator based on the unconditional distributional PT assumption that ignores covariates. Finally, $F_{ Y_{t}(0)|d_r=1}^{-1}$ represents the true population parameter, whereas `Bias' and `RMSE' stand for average (simulated) bias, and root-mean-squared, respectively.}}
\end{table}

\begin{table}[!htbp]
\center
\small
\resizebox{0.8\textwidth}{!}{
\caption{Monte Carlo Results, DGP 3, $\tau=.5$. Setup with $T=R=4$.}
\label{MC_dgp3_appendix}
\begin{tabular}{l*{6}{c}}
\toprule
\vspace*{0.0002cm}	\\																								
	&	&	&		 \multicolumn{2}{c}{$n=100$} 				&	 \multicolumn{2}{c}{$n=1,000$} 			\\
			\cmidrule(rl){4-5}						\cmidrule(rl){6-7}				
\vspace*{0.0025cm}	\\												
		&	&	$F_{ Y_{t}(0)|d_r=1}^{-1}$ 	&	Bias	&	Root MSE 	&	Bias	&	Root MSE 	\\
\multirow{6}{4em}{\textbf{Unc}}	&	$F_{Y_{2}(0)|d_2=1}^{-1}$	&	4.839	&	0.244	&	0.583	&	0.085	&	0.18	\\
	&	$F_{Y_{3}(0)|d_2=1}^{-1}$	&	5.841	&	0.239	&	0.571	&	0.083	&	0.179	\\
	&	$F_{Y_{4}(0)|d_2=1}^{-1}$	&	6.867	&	0.2	&	0.55	&	0.061	&	0.171	\\
	&	$F_{Y_{3}(0)|d_3=1}^{-1}$	&	6.837	&	0.239	&	0.561	&	0.092	&	0.178	\\
	&	$F_{Y_{4}(0)|d_3=1}^{-1}$	&	7.869	&	0.206	&	0.544	&	0.064	&	0.163	\\
	&	$F_{Y_{4}(0)|d_4=1}^{-1}$	&	8.867	&	0.206	&	0.527	&	0.065	&	0.162	\\
 \vspace*{0.0025cm}	\\												
\multirow{6}{4em}{\textbf{Cond}}	&	$F_{Y_{2}(0)|d_2=1}^{-1}$	&	4.839	&	0.203	&	0.59	&	0.04	&	0.169	\\
	&	$F_{Y_{3}(0)|d_2=1}^{-1}$	&	5.841	&	0.195	&	0.576	&	0.036	&	0.168	\\
	&	$F_{Y_{4}(0)|d_2=1}^{-1}$	&	6.867	&	0.16	&	0.56	&	0.015	&	0.167	\\
	&	$F_{Y_{3}(0)|d_3=1}^{-1}$	&	6.837	&	0.191	&	0.573	&	0.041	&	0.168	\\
	&	$F_{Y_{4}(0)|d_3=1}^{-1}$	&	7.869	&	0.16	&	0.565	&	0.012	&	0.163	\\
	&	$F_{Y_{4}(0)|d_4=1}^{-1}$	&	8.867	&	0.157	&	0.545	&	0.017	&	0.158	\\
\bottomrule
\end{tabular}
\floatfoot{\footnotesize{\textbf{Notes}: This table reports the Monte Carlo results for all the parameters of the DGP in \hyperref[DGP3]{Eq. (\ref*{DGP3})} in the setup with $T=R=4$ for the quantile $\tau=.5$. Each Monte Carlo simulation uses $2,000$ bootstrap replications. Rows labeled `UNC' use the estimator based on the unconditional distributional PT assumption that ignores covariates. Finally, $F_{ Y_{t}(0)|d_r=1}^{-1}$ represents the true population parameter, whereas `Bias' and `RMSE' stand for average (simulated) bias, and root-mean-squared, respectively.}}}
\end{table}

\begin{table}[!htbp]
\resizebox{\textwidth}{!}{
\caption{\footnotesize Monte Carlo Results, DGP2, for $F_{Y_{2}(0)|d_2=1, \rho=0}^{-1}(\tau)$ . Setup with $T=R=4$ and non-linear trends.}
\label{montecarlo_nonlinear_trends}
\resizebox{\textwidth}{!}{
\begin{tabular}{l*{10}{c}}
\toprule
\small
\vspace*{0.0002cm}	\\																								
\vspace*{0.0002cm}	\\																								
				&		 & \multicolumn{3}{c}{\textbf{0.25}} 					&\multicolumn{3}{c}{\textbf{0.50}} 				& \multicolumn{3}{c}{\textbf{0.75}} 			\\							
	\cmidrule(rl){3-5} \cmidrule(rl){6-8} \cmidrule(rl){9-11} 																								
\vspace*{0.0025cm}	\\																								
		&	&	$F_{ Y_{t}(0)|d_r=1}^{-1}$ 	&	Bias	&	Root MSE 	&		$F_{ Y_{t}(0)|d_r=1}^{-1}$ 	&	Bias	&	Root MSE 	&		$F_{ Y_{t}(0)|d_r=1}^{-1}$ 	&	Bias	&	Root MSE 	\\
  \vspace*{0.0000001cm}	\\																								
\cline{2-11}																									
  \vspace*{0.0025cm}	\\																								
\multirow{2}{4em}{$n=100$}	&	\textbf{Unc}	&	6.832	&	0.331	&	0.684	&	8	&	0.344	&	0.672	&	9.168	&	0.392	&	0.715	\\
	                    &	\textbf{Cond}	&	6.832	&	0.082	&	0.526	&	8	&	0.09	&	0.508	&	9.168	&	0.152	&	0.553	\\
\vspace*{0.005cm}																									\\
\multirow{2}{4em}{$n=1000$}	&	\textbf{Unc}	&	6.832	&	0.256	&	0.315	&	8	&	0.252	&	0.305	&	9.168	&	0.25	&	0.31	\\
	                    &	\textbf{Cond}	&	6.832	&	0.017	&	0.158	&	8	&	0.013	&	0.149	&	9.168	&	0.011	&	0.156	\\
\bottomrule
\end{tabular}}
\floatfoot{\footnotesize{\textbf{Notes}: This table reports the Monte Carlo results for the estimator of the parameter $F_{Y_{2}(0)|d_2=1, \rho=0}^{-1}(\tau)$ of the DGP in \hyperref[DGP2]{Eq. (\ref*{DGP2})} in the setup with $T=R=4$ and with a quadratic trend (i.e., $\alpha_t=t+t^2$). Results are reported for the quantiles $.25$, $.50$, $.75$. Each Monte Carlo simulation uses $2,000$ bootstrap replications. Rows labeled `UNC' use the estimator based on the unconditional distributional PT assumption that ignores covariates. Finally, $F_{ Y_{t}(0)|d_r=1}^{-1}$ represents the true population parameter, whereas `Bias' and `RMSE' stand for average (simulated) bias, and root-mean-squared, respectively.}}}
\end{table}